\documentclass[twocolumn]{aastex61}

\usepackage{soul}

\received{}
\revised{}
\accepted{June 22, 2018}
\shorttitle{Supersoft X-ray sources identified with
 Be binaries in the Magellanic Clouds}
\shortauthors{Cracco et al. 2018}
\begin{document}

\title{Supersoft X-ray sources identified with
 Be binaries in the Magellanic Clouds\footnote{Based on observations made with the Southern African Large Telescope (SALT).}}

\author{Valentina Cracco}
\affiliation{Department of Physics and Astronomy, Padova University,
vicolo Osservatorio, 3 - 35122 Padova, Italy }

\author{Marina Orio}
\affiliation{Department of Astronomy, University of Wisconsin
475 N. Charter Str.  Madison, WI 53706}
\affiliation{INAF - Astronomical Observatory Padova, vicolo dell'Osservatorio 5, 35122 Padova, Italy}

\author{Stefano Ciroi}
\affiliation{Department of Physics and Astronomy, Padova University,
vicolo Osservatorio, 3 - 35122 Padova, Italy }

\author{Jay Gallagher}
\affiliation{Department of Astronomy, University of Wisconsin
475 N. Charter Str.  Madison, WI 53706}

\author{Ralf Kotulla}
\affiliation{Department of Astronomy, University of Wisconsin
475 N. Charter Str.  Madison, WI 53706}

\author{Encarni Romero-Colmenero}
\affiliation{South African Astronomical Observatory/Southern African Large Telescope, P.O. Box 9, Observatory, 7935, South Africa}



\begin{abstract}
We investigated four luminous supersoft X-ray sources (SSS) in the Magellanic Clouds suspected to have optical counterparts of Be spectral type. If the origin of the X-rays is in a very hot atmosphere heated by hydrogen burning in accreted envelopes of white dwarfs (WDs), like in the majority of SSS, these
objects are close binaries, with very massive WD primaries. Using the South African Large Telescope (SALT), we obtained  the first optical spectra of the proposed optical counterparts of two candidate Be stars associated with SUZAKU J0105-72 and XMMU J010147.5-715550, respectively a transient and a recurrent SSS, and
confirmed the proposed Be classification and Small Magellanic Clouds membership. We also obtained new optical spectra of two other  Be stars
 proposed as optical counterparts of the transient SSS XMMU J052016.0-692505 and MAXI-J0158-744. The optical spectra with double peaked emission line profiles, are typical of Be stars and present characteristics similar to many high mass X-ray binaries with excretion disks, truncated by the tidal interaction with a compact object. The presence of a massive WD  that sporadically ignites nuclear burning, accreting only at certain orbital or evolutionary phases, explains the supersoft X-ray flares. We measured equivalent widths and distances between lines' peaks, and investigated the variability of the prominent emission lines' profiles. The excretion disks seem to be small in size, and  are likely to be differentially rotating. We discuss possible future observations and the relevance of these objects as a new class of type Ia supernovae progenitors.
\end{abstract}

\keywords{stars: emission-line,Be --- line: profiles --- techniques: spectroscopic --- Magellanic Clouds}



\section{Introduction} \label{sec:intro}
 Very luminous supersoft X-ray sources (SSS) have been observed since the end of the '70ies with the Einstein satellite, but they still pose  unsolved riddles.
 More than half of the SSS are transient sources,  and a majority turn out to be post-outburst novae,
 in which the WD keeps on burning hydrogen for a period of time ranging from a week to  years after the outburst, with an atmospheric temperature of up to a million K. Novae are observed as SSS in the Galaxy and
 the Magellanic Clouds, and now routinely even in M31 \citep[for lists and reviews of novae in the Galaxy and in the Local Group see among others][]{Orio2010, Orio2012, Henze2011, Henze2014, Osborne2015}.
 However, the nature of numerous SSS  is not yet understood, and may hold the key to outstanding astrophysical problems. Because of the large intrinsic luminosity, these sources are observed in the direction of external galaxies, in the Local Group and beyond, up to a distance of 15 Mpc
 \citep[e.g.][]{DiStefano2004, DiStefano2003}, in regions of the sky affected by low absorption. Many non-nova
SSS are likely to be in close binaries hosting the hottest, most massive accreting and hydrogen burning WDs, possibly on the verge of type Ia supernova explosions \citep[SNe Ia; see for instance][]{vandenHeuvel1992}. The phenomenology of SSS has an overlap with accreting black holes \citep[see, e.g.,][]{Liu2008} and the most luminous SSS may host, instead of WDs, massive stellar black holes accreting at super-Eddington luminosity.

A group of SSS appears particularly interesting, potentially explaining the dependence of the SNe Ia rate star formation rate \citep[e.g.][]{Sullivan2006} even  without invoking double degenerate systems: the SSS
that are associated with massive stars, proposed to be Be type stars \citep[see review by][]{Orio2013}.
In addition to four SSS whose new observation we describe in this paper,
 there is a fifth SSS in the Magellanic Clouds: the transient source
RX J0527.8-6954 \citep[][]{Greiner1991, Greiner1996}),
 whose X-ray light curve is like that of a classical nova, has been positively identified
with a Be optical counterpart \citep[][]{Oliveira2010}.
 This source is in a very crowded field and other possible counterparts 
were previously proposed, but the identification by \citet[][]{Oliveira2010}
with a B5e V star with subarsecond and possibly bipolar H$\alpha$ emission appears very convincing. This object also shows weak [O III] emission.
 Moreover, among almost a hundred SSS in M31, in a statistically significant fraction
 of them have only massive young stars as detectable
 candidate optical counterparts \citep[][]{Orio2010}.

Be stars with WD companions have not been detected yet, but there is reasonable circumstantial evidence that the rapidly rotating B star Regulus, for instance, hosts a WD companion \citep{Gies2008}. The mass range of B stars spans from a little above 2 M$_\odot$ to 20 M$_\odot$, so most of these stars end as WDs.
Thus, the binary may have progenitor components in this
 range, and does not require an unusual evolutionary path
 with large mass loss or mass transfer to end
 as a WD+Be system. However,
 mass transfer does also allow for unusual configurations, in which the initial secondary, or less massive star, ends its life as a WD before the more massive companion. \citet{Matson2015}  describe observational evidence obtained for KOI-81, a system in which the original donor star was completely stripped and the secondary (a B star) was spun up. \citet{Wang2018} discovered many cases of Be stars with hot, stripped-down sub-dwarfs detected in the ultraviolet.
 
\citet[][]{Waters1989}
 predicted a large fraction of Be+WD systems and
\citet[][]{Raguzova2001} calculated that 70\% of Be systems have a WD companion,
 with most orbital periods less than a year.
 Such systems are not long lived as SSS, but altogether last longer than neutron star+Be star systems, of which many are known \citep[about 90 are known in the SMC, and 33 in the LMC][]{Antoniou2016, Antoniou2014}.
\citet[][]{McSwain2005} found that more than 70\% of Be binaries are spun up in the course of evolution and are expected to end as close binaries.
 Only one Be star with a black hole companion has been discovered \citep[][]{Casares2014}, but 
it does not emit luminous supersoft X-rays;
 it is instead a rather hard X-ray source of low X-ray luminosity (1.6 $\times$ 10$^{-7}$ times the Eddington luminosity).

We obtained low resolution spectra of the proposed optical counterparts
 of four SSS in the Magellanic Clouds with the Southern African Large Telescope \citep[SALT,][]{2006SPIE.6267E..0ZB} 10-m telescope, and we show in this paper that all four have characteristics of Be stars.
 Two of the sources had already been observed spectroscopically, so we could assess
 whether the spectra evolved.
In Section \ref{sec:sample} we present the targets,
 the observations and reduction procedures are described in Section \ref{sec:obs}, in Section \ref{sec:analysis} we show the analysis and we discuss the results in Section \ref{sec:discussion}. A summary and conclusions are presented in Section \ref{sec:summary}.

\section{The four targets}\label{sec:sample}
None of our targets were persistent X-ray sources. Three of them were transients, but have
 not been observed often enough to rule out recurrence. We know
 that one, XMMU-J010147.5-715550, is a variable, recurrent SSS.

\subsection{Previously proposed identifications without optical spectra}
\citet[][]{Sturm2012} discovered XMMU J010147.5-715550, a recurrent SSS  in the SMC,
 with several detections of different significance
 in 13 out of 28 exposures done with {\sl XMM-Newton}
between 2000 and 2010. It was
measured at several $\sigma$ detection level in 2000-2001,
 and there were some marginal detections (1-2 $\sigma$ level) until
 2011.  We found that this SSS was observed again 
 with {\sl XMM-Newton} 20 times, in April and in October and/or November of each year
 from 2011 to 2017, but it  was never detected again, with upper limits
 of  about 10$^{34}$ erg s$^{-1}$ for SMC distance.
 In 2000-2001, the unabsorbed flux was close to 10$^{-14}$ erg cm$^{-2}$ s$^{-1}$,
 but due to uncertainties in fitting the
 spectrum and especially on the column density,
 only a lower limit to the absolute luminosity at SMC distance of
 few times 10$^{34}$ erg s$^{-1}$ was derived by \citet[][]{Sturm2012}.
 Because this is two orders of magnitude below the lower
 luminosity range of other known SSS,
 the SSS classification is not quite certain. However, XMMU J010147.5-715550 
 was a very soft X-ray source. 
  \citet[][]{Sturm2012} found only a Galactic foreground star and a likely B star at
 V=14.30$\pm$0.04 in the SMC in the spatial error box. This star was classified as
  O7IIIe-B0Ie  \citep[see][and references therein]{Sturm2012}, where the
classification as an emission-line star  is made because
 of the spatial coincidence with an H$\alpha$
point source in narrow band photometric images.
An O star, as in a recent classification by the OGLE team by
\citet{Kourniotis2014}, is more likely to have a black hole companion. A WD companion would imply that the primary has undergone extreme mass transfer, but has formed a carbon-oxygen core that, stripped of all the envelope, has contracted forming a WD. In fact the density of a carbon-oxygen or oxygen-neon WD  is necessary to ignite hydrogen CNO burning in a shell in degenerate conditions. However, the average magnitude V=14.4 is in the range of luminosity of a B star.
 This object is classified as optically variable by the OGLE team, without a known periodicity,
but rather with variability on different time scales,
an amplitude of 0.2 mag in the I filter, and overall increase in luminosity over
 about 500 days \citep[][]{Kourniotis2014}.
\citet[][]{Sturm2012} found a tentative periodicity of 1264$\pm$2 days in the I-band
 OGLE III light curve and discussed why the X-ray properties indicate  an SMC intrinsic source;
 with two different methods they estimated that the probability of coincidence of an
emission line star with an X-ray  source in the SMC is only of order 0.6-0.7\%.
These authors thus suggested that 
 the compact object in XMMU J010147.5-715550 is a WD, 
 accreting and igniting hydrogen at orbital phases that bring it 
close to the Be star; this causes the SSS emission to be recurrent rather than stable.
 They suggest that the putative WD accretes only when an excretion disk is present, and such disks
 around Be stars are known to have a transient-recurrent nature.

Suzaku J0105-72 was a transient SSS observed  in a 2005 March
 Suzaku observation of a supernova remnant, observed
 many times, and it was luminous only
 in one out of 16 exposures done with {\sl Suzaku} between
 2003 and 2007. However, the SSS was
still marginally detected with Chandra ACIS-S in a deeper
 image on 2008 January 27,
 with a flux about three orders of magnitude lower than in the {\sl Suzaku observations}.
 There were no additional detections, in 29 {\sl XMM-Newton} observations
 done before 2007 March and 22 done after  2008 January (not evenly spaced,
 but concentrated
 in few days three times a year),
with an upper limit to that was three orders of magnitudes lower than the
flux in the detection image \citep[][]{Takei2008}.
 The same field was also imaged 18 times with the {\sl ROSAT} PSPC and HRI between 1993 and 1998,
 yielding no detections with flux upper limits about two orders of magnitude lower than the
 the average of the Suzaku
 observation. At SMC distance,
 the X-ray luminosity during the observation was in fact about 2 $\times$ 10$^{37}$
 erg s$^{-1}$, assuming the best fit parameters N(H)=4.9 $\times$ 10$^{20}$ erg s $^{-1}$ and a
72 eV blackbody temperature, but actually it 
 steadily decreased during 24 ks of exposure. Although the source was at the
 edge of the field, an accurate position was derived with ray-tracing methods and
the most likely optical counterpart is a star  of spectral type B, measured at B=14.64
\citep{Evans2004}. While  \citet[][]{Evans2004} classified the optical star  as a B0 IV
spectral type,
recently
\citet[][]{Lamb2016} re-classified it as an O9.5 III/V e star, however also
 in this case, the optical magnitude (corresponding to
absolute magnitude $V=-3.4$ at SMC distance),
is rather in the range of a Be star rather than an extremely rare Oe star.
 The OGLE team classifies it as a variable object with a tentative period
 of 21.823 days \citep{Kourniotis2014}.

\subsection{Proposed identification with published optical spectroscopy}
XMMU J052016.0-692505 was observed as an SSS in {\sl XMM-Newton} observations done
 on 2004 January 17 \citep[][]{Kahabka2006}. Assuming that
 there is an amount of circumstellar absorption, 
 \citet{Kahabka2006} 
 fitted the spectrum with a column  density value N(H)=2.8$\times 10^{21}$ cm$^{-2}$ (the column density along the line of sight is N(H)=4.7$\times 10^{20}$ cm$^{-2}$), a blackbody at 33 eV and unabsorbed luminosity 10$^{36}$ erg s$^{-1}$ \citep[however 
 the fit had large uncertainties due to the data quality,
 see][]{Kahabka2006}. The source was marginally detected in several {\sl ROSAT}
 observations in 1997. \citet[][]{Kahabka2006} suggested the optical identification
with a luminous blue star at V=15.45$\pm$0.05, tentatively classified as a B star
 and known to be variable with a possible periodicity
 of  510$\pm$20 days and/or 1040$\pm$70
 days in MACHO and OGLE data \citep[see results and references in][]{Kahabka2006}.
 \citet[][]{Kahabka2006} suggested that this is
 a Be/WD binary, in which the WD accretes hydrogen from the excretion disk and burns it
 on the surface.  The optical spectrum published by \citet[][]{Kahabka2006}
 shows prominent H$\alpha$ and H$\beta$ emission lines  and no \ion{He}{2} at 4686 \AA.

MAXI-J0158-744 differs from other SSS because it was a very brief and luminous transient event, and it was not  supersoft from the beginning.  It was initially detected with {\sl MAXI}
 on 2011 November 11  as an extremely luminous flare, with absolute X-ray
luminosity close to 10$^{40}$ erg s$^{-1}$  assuming  only the interstellar absorption due to
 the column density along the line of sight, N(H)=4 $\times$ 10$^{20}$ cm$^{-2}$. In the following
 two weeks, it was observed as an SSS with {\sl Swift} while the luminosity decreased rapidly
\citep[][]{Li2012, Morii2013}. This behavior resembles that of the ``state changing'' ultra luminous X-ray sources in
nearby galaxies  outside the Local Group \citep[][]{Liu2011}. The only optical 
counterpart in the spatial error bar is a luminous star, which
turned  out to be a  Be star
 with magnitude I=14.82$\pm$0.01 at quiescence,
 with prominent emissions: Balmer hydrogen lines, several \ion{He}{1} lines and a  relatively weak line of \ion{He}{2} at 4686 \AA\ \citep[][]{Li2012}.
The source had become more optically luminous by 0.53 magnitudes in an observation
a day or two after the initial X-ray flare, and it was observed to return to quiescent optical
luminosity.
Both \citet[][]{Li2012} and \citet[][]{Morii2013} suggested it must have been a nova event in a Be+WD system.

\section{The optical spectra: observations and data reduction}\label{sec:obs}

We observed the four proposed optical counterparts described above between 2016
September and October with the
Robert Stobie Spectrograph \citep[RSS,][]{2003SPIE.4841.1463B,2003SPIE.4841.1634K} and between 2017 June and October with the echelle
 High Resolution Spectrograph \citep[HRS,][]{2010SPIE.7735E..4FB,2012SPIE.8446E..0AB,2014SPIE.9147E..6TC}  mounted at the (SALT).

The observations are summarised in Table~\ref{tab2}. XMMU J010147.5-715550, SUZAKU J0105-72 and XMMU-J052016-692505 were observed with the PG0900 grating in long slit mode with a slit field of view (f.o.v.) of 1$^{\prime\prime}\times 8^{\prime}$. The resulting spectral resolution is about $R = 1100$, with a dispersion of about 0.97 \AA\ pixel$^{-1}$ and the wavelength range is about 4060--7130 \AA. MAXI-J0158-744 and XMMU-J052016-692505 were observed with the PG2300 grating in long slit mode with the same slit.
The resulting spectral resolution is about $R = 2900$, with a dispersion of 0.32 \AA\ pixel$^{-1}$. The wavelength range is 4440-5460 \AA.
All the targets were observed with the HRS echelle spectrograph with its
 low resolution grating,
 which allows to observe the 3700-8870 \AA\ spectral range with spectral resolution $R=15000$ and a dispersion between 0.024 and 0.045 \AA\ pixel$^{-1}$.

 The extraction, reduction and analysis of spectra were performed with Image Reduction and Analysis Facility (IRAF -- version 2.16.1\footnote{IRAF is distributed by the National Optical Astronomy Observatories, which are operated by the Association of Universities for Research in Astronomy, Inc., under cooperative agreement with the National Science Foundation.}).
 The spectra provided by SALT are already bias-subtracted. We extracted the RSS spectra with the {\sc apall} task, wavelength-calibrated and flux-calibrated using the spectrophotometric standard stars LTT6248, EG274, LTT7379 and HILT600 and the usual IRAF reduction tasks. The standard stars were observed
with the widest slit (width = 4$^{\prime\prime}$) and in the same position along the slit as the targets. We reduced the echelle spectra using the {\sc echelle} package in {\sc iraf}. We performed the tracing using the flat-field frames, which have higher signal-to-noise ratio (S/N) than the objects. The objects were corrected for cosmic rays and extracted. Then, wavelength calibration was applied using a thorium-argon lamp and the sky contribution for each aperture was subtracted. The spectra were flux calibrated using the HR7596, HR5501, HR14943 standard stars and the usual {\sc standard} and {\sc sensfunc} tasks.

 Absolute flux calibration cannot be achieved with the SALT telescope because it has a variable pupil and the illuminating beam changes during the observations, therefore we only obtained relative flux calibration.
 The spectra shown in this work are either flux-normalized spectra, using the flux values
 at $\lambda=5500$ \AA\ (RSS with PG0900) or $\lambda=4950$ \AA\ (RSS with PG2300),
 or flux-calibrated spectra (HRS), where however the calibration is relative (not
 absolute).
%%%%
%%%%
\begin{table*}
\begin{center}
\caption{The SALT observations}
{\tiny
\begin{tabular}{cccccccccc}
\hline
Object & Date & Instrument & Grating & Grating angle & Spectral range & $R$ & $\delta \lambda$ & T$_{\rm exp}$\\
   & & & & ($^{\circ}$) & (\AA) & ($\lambda/\Delta \lambda$) & (\AA\ px$^{-1}$) & (s) \\
\hline
XMMU-J010147.5-715550 & 2016-09-07 & RSS & PG0900 & 14.75 & 4063--7113 & 1100 & 0.96 & 900 \\
                      & 2016-09-29 & RSS & PG0900 & 14.75 & 4062--7137 & 1100 & 0.96 & 900 \\
                      & 2017-06-17 & HRS & LR & -- & 3702--8870 & 15000 & 0.024--0.045 & 1700 \\
                      & 2017-07-30 & HRS & LR & -- & 3702--8870 & 15000 & 0.024--0.045 & 1700 \\
                      & 2017-10-26 & HRS & LR & -- & 3702--8870 & 15000 & 0.024--0.045 & 1400 \\
SUZAKU-J0105-72       & 2016-09-20 & RSS & PG0900 & 14.75 & 4061--7132 & 1100 & 0.97 & 900 \\
                      & 2016-10-19 & RSS & PG0900 & 14.75 & 4063--7130 & 1100 & 0.97 & 700 \\
                      & 2017-07-14 & HRS & LR & -- & 3702--8870 & 15000 & 0.024--0.045 & 1900 \\
                      & 2017-08-26 & HRS & LR & -- & 3702--8870 & 15000 & 0.024--0.045 & 1900 \\
                      & 2017-10-26 & HRS & LR & -- & 3702--8870 & 15000 & 0.024--0.045 & 1600 \\
XMMU-J052016-692505   & 2016-09-20 & RSS & PG2300 & 35.00 & 4443--5463 & 2900 & 0.32 & 960 \\
                      & 2016-10-20 & RSS & PG2300 & 35.00 & 4442--5462 & 2900 & 0.32 & 960+669 \\
                      & 2016-10-25 & RSS & PG0900 & 14.75 & 4062--7131 & 1100 & 0.97 & 870 \\
                      & 2017-09-17 & HRS & LR & -- & 3702--8870 & 15000 & 0.024--0.045 & 1575 \\
                      & 2017-10-22 & HRS & LR & -- & 3702--8870 & 15000 & 0.024--0.045 & 1575 \\
                      & 2017-10-29 & HRS & LR & -- & 3702--8870 & 15000 & 0.024--0.045 & 1400 \\
MAXI-J0158-744        & 2016-10-10 & RSS & PG2300 & 35.00 & 4443--5463 & 2900 & 0.32 & 900 \\
                      & 2016-10-23 & RSS & PG2300 & 35.00 & 4443--5462 & 2900  & 0.32 & 900 \\
                      & 2017-08-13 & HRS & LR & -- & 3702--8870 & 15000 & 0.024--0.045 & 1750 \\
                      & 2017-09-17 & HRS & LR & -- & 3702--8870 & 15000 & 0.024--0.045 & 1750 \\
                      & 2017-10-26 & HRS & LR & -- & 3702--8870 & 15000 & 0.024--0.045 & 1350 \\
\hline
\label{tab2}
\end{tabular}
}
\end{center}
\end{table*}

\section{Data analysis}\label{sec:analysis}
In the RSS spectra taken with the PG2300 grating and in the HRS spectra we
 observed double peaked line profiles, which are typical of the rotating
 disks of the Be stars \citep[see, among others, recent work by][]{2018AJ....155...53L}.
In the RSS spectra taken with PG0900 grating,
 we could not resolve two peaks even if they were present, due to low resolution. 
 In all RSS spectra, we measured the velocity of the centroid of the lines by means of the IRAF task {\sc emsao}. Because the line profiles are complex 
 in the HRS spectra, we could not apply {\sc emsao}, and
 we estimated the radial velocity by measuring the position of both the blue/violet (V) and red (R) peaks, then we took the average of the two peaks.
We analyzed all the spectra with the IRAF task {\sc splot} to measure the
 equivalent width (EW) and to fit the lines. We
 fitted the emission lines with one or  more Gaussian functions,
 and the underlying continuum with a straight line. To estimate the errors, we chose ten
 different continuum levels to measure EWs and we fitted the profiles ten times,
 thus  obtaining the mean and root mean square (rms) values for both EW, FWHM and
 centroid positions. For double-peaked profiles, we also measured V/R, defined as
 the ratio of the V and the R peak intensity
\citep[to compare our work with that of][]{Dachs+92}.
 The mean EW and rms values are given
in Table~\ref{tab3}, while kinematics data and V/R are in Table~\ref{tab4}.

\begin{figure*}
\includegraphics[width=0.33\textwidth]{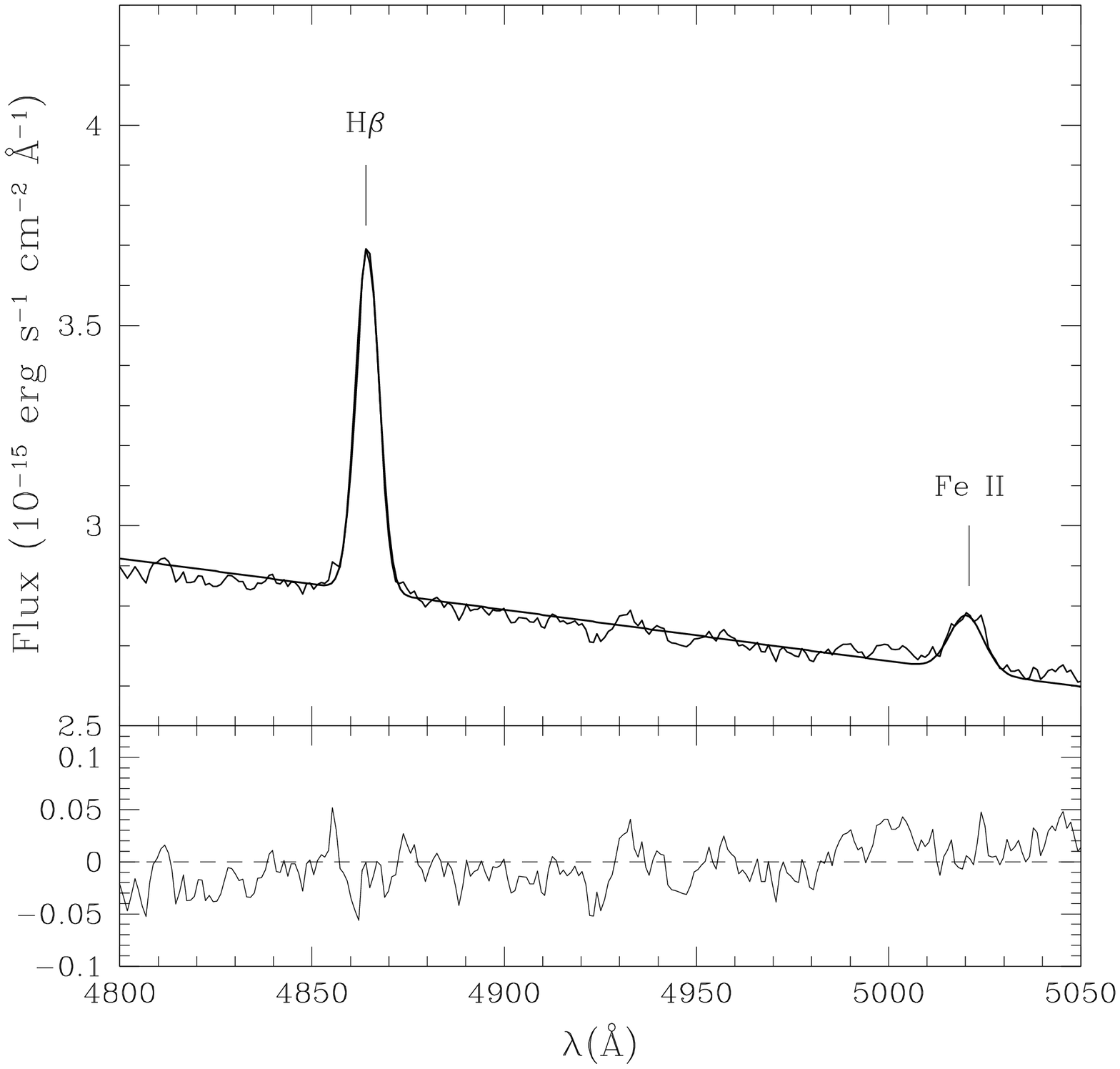}
\includegraphics[width=0.33\textwidth]{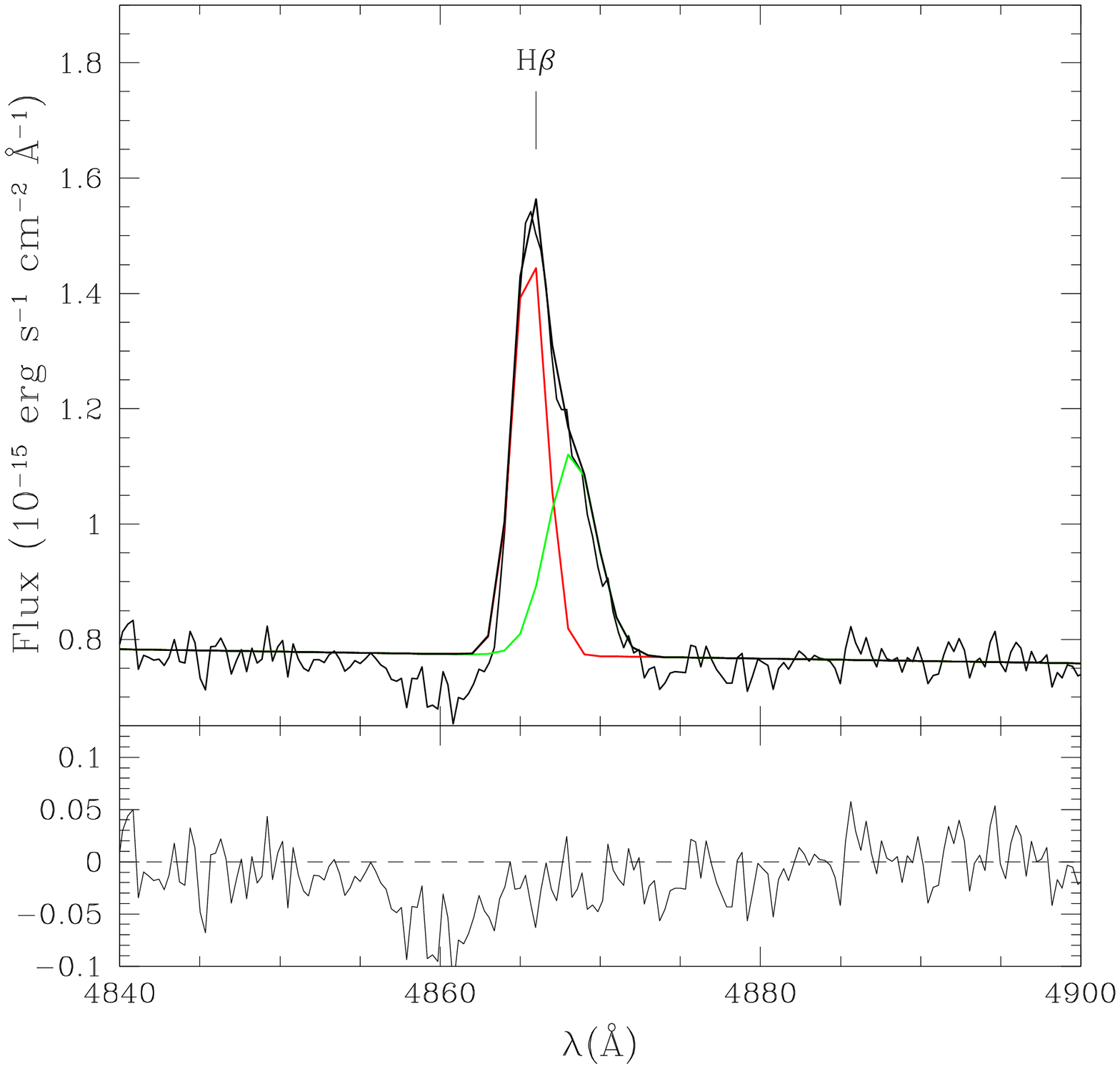}
\includegraphics[width=0.33\textwidth]{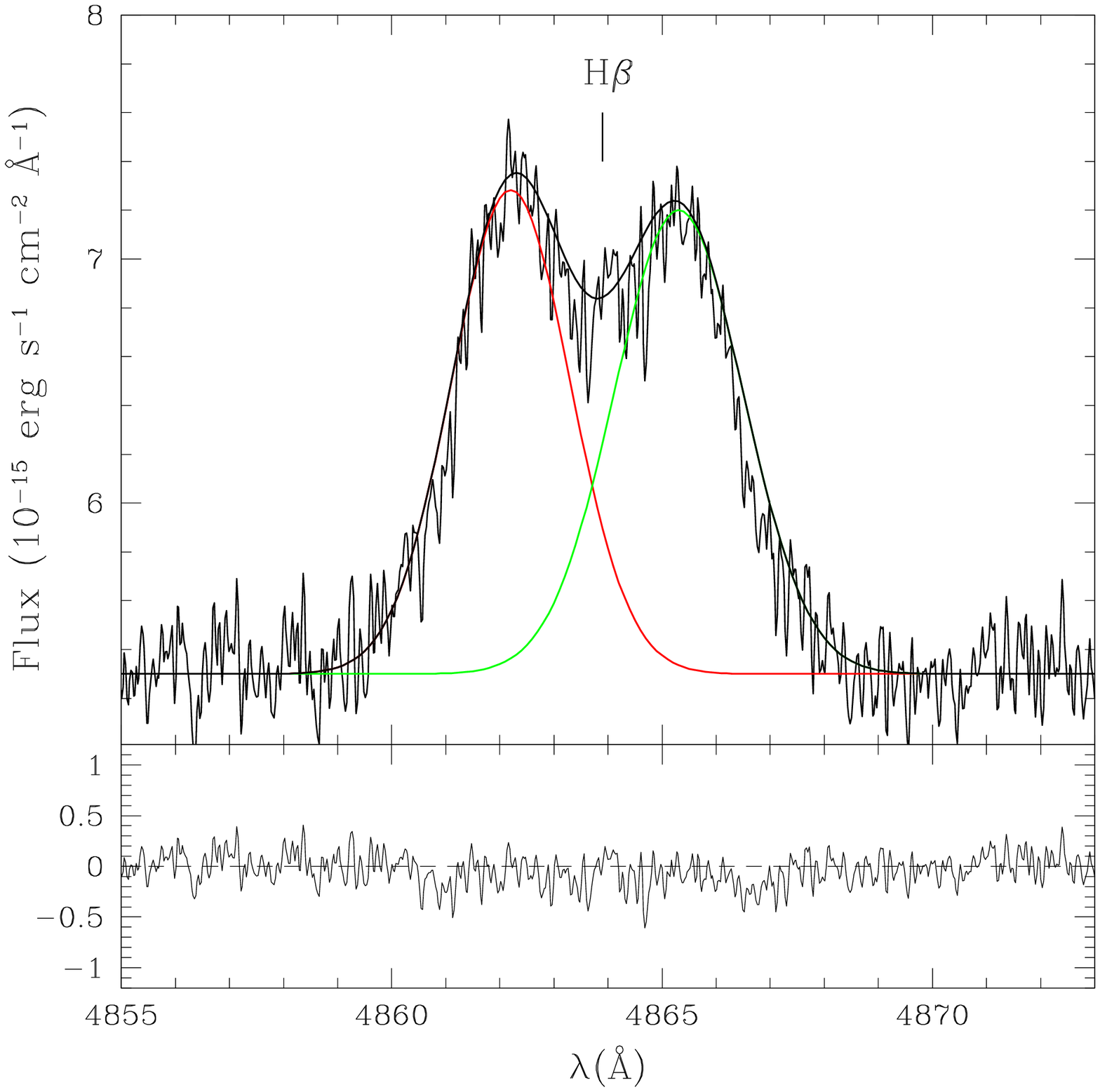}
\caption{In these plots, the continuum was fitted with a straight line and the emission lines with one or more Gaussians. The panel on the left shows how we fitted the emission lines in the PG0900  spectrum  (here we show H$\beta$ and \ion{Fe}{2}\,$\lambda5018$). The central panel shows an example of how we fitted the H$\beta$ line in the PG2300 grating spectra, with two Gaussian functions,
 plotted in red and green,
 respectively.
 Finally, the third panel shows the fit we were able to do for the lines measured with the HRS, with two Gaussians (here we show an H$\beta$ line). The residuals of the fits are shown below each plot.}
\label{fig1}
\end{figure*}

We fitted the lines that were clearly split, with  an R and a V peak,
 (PG 2300 RSS and HRS spectra)
 with two Gaussian functions to measure the velocity difference ($\Delta v$) of the V
 and R peak.
 Presumably, assuming the lines arise in a disk,  these velocity differences represent $2 v_d \sin i$, where $v_d$ is the Keplerian velocity of a characteristic radius in the outer disk where the line flux is a maximum, and $i$ is the disk inclination to the line of sight. When two peaks were not clearly measurable, for all  PG0900
 RSS spectra and for some lines in the HRS spectra, we assumed that an estimate for
 $\Delta v$ is given by the velocity dispersion value $\sigma = 0.424 \times {\rm FWHM}$.
 The October spectra of XMMU J052016-692505 obtained with HRS show
 complex lines in which  three peaks are 
 detected, so we measured the FWHM directly on the line
 profiles and we derived $\sigma$ from the FWHM. 
In Fig.~\ref{fig1} we show examples of how we fitted lines observed with the different gratings or spectrographs. The wavelength-calibration errors were calculated by assuming a mean error of 20\%\ of the dispersion value: the mean errors  are 11, 4 and 2 km s$^{-1}$, for the PG0900, PG2300 and HRS data, respectively.
%%%%

\subsection{Analysis of the spectra}
\begin{figure}
\begin{center}
\includegraphics[width=\columnwidth]{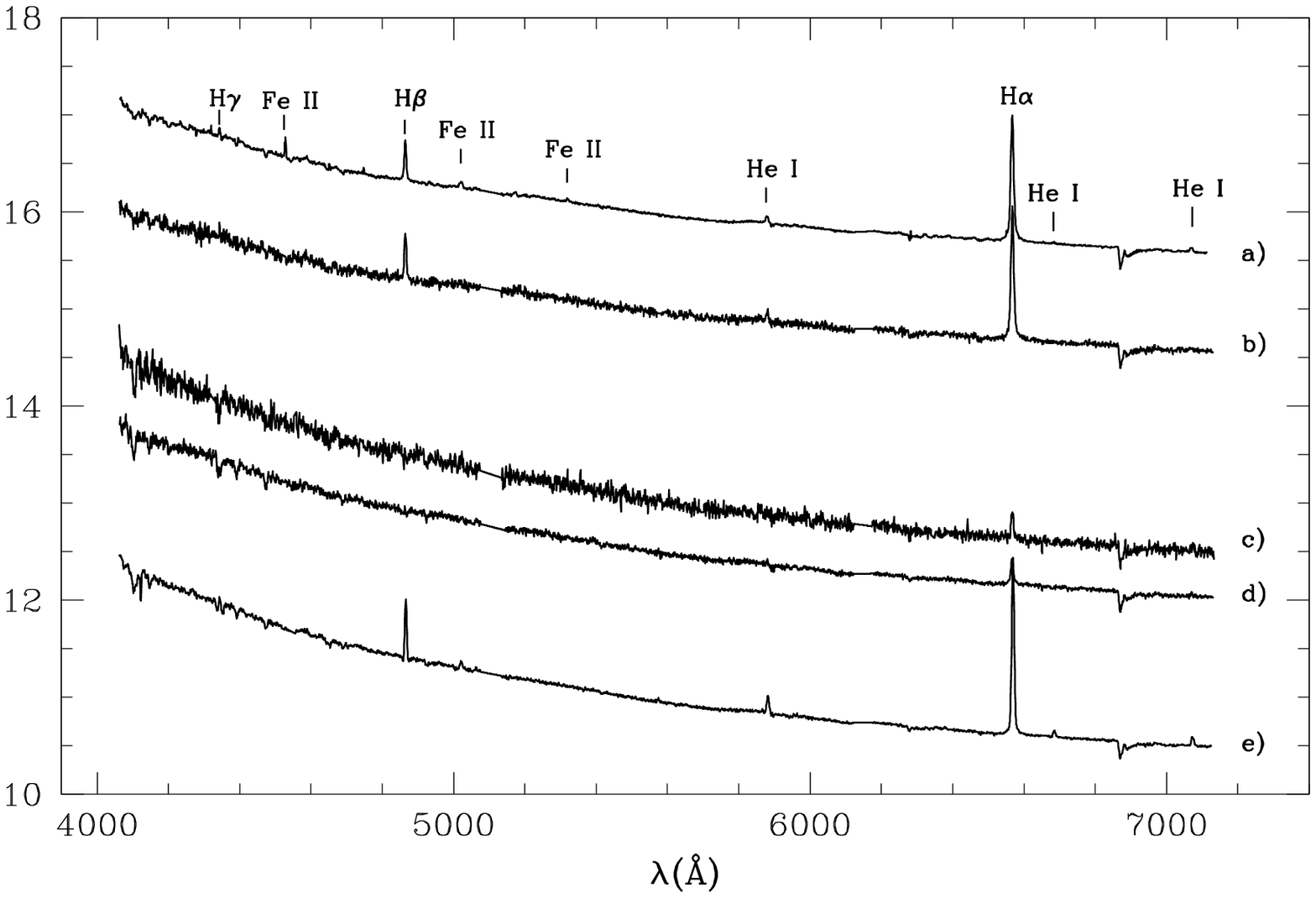}
\includegraphics[width=\columnwidth]{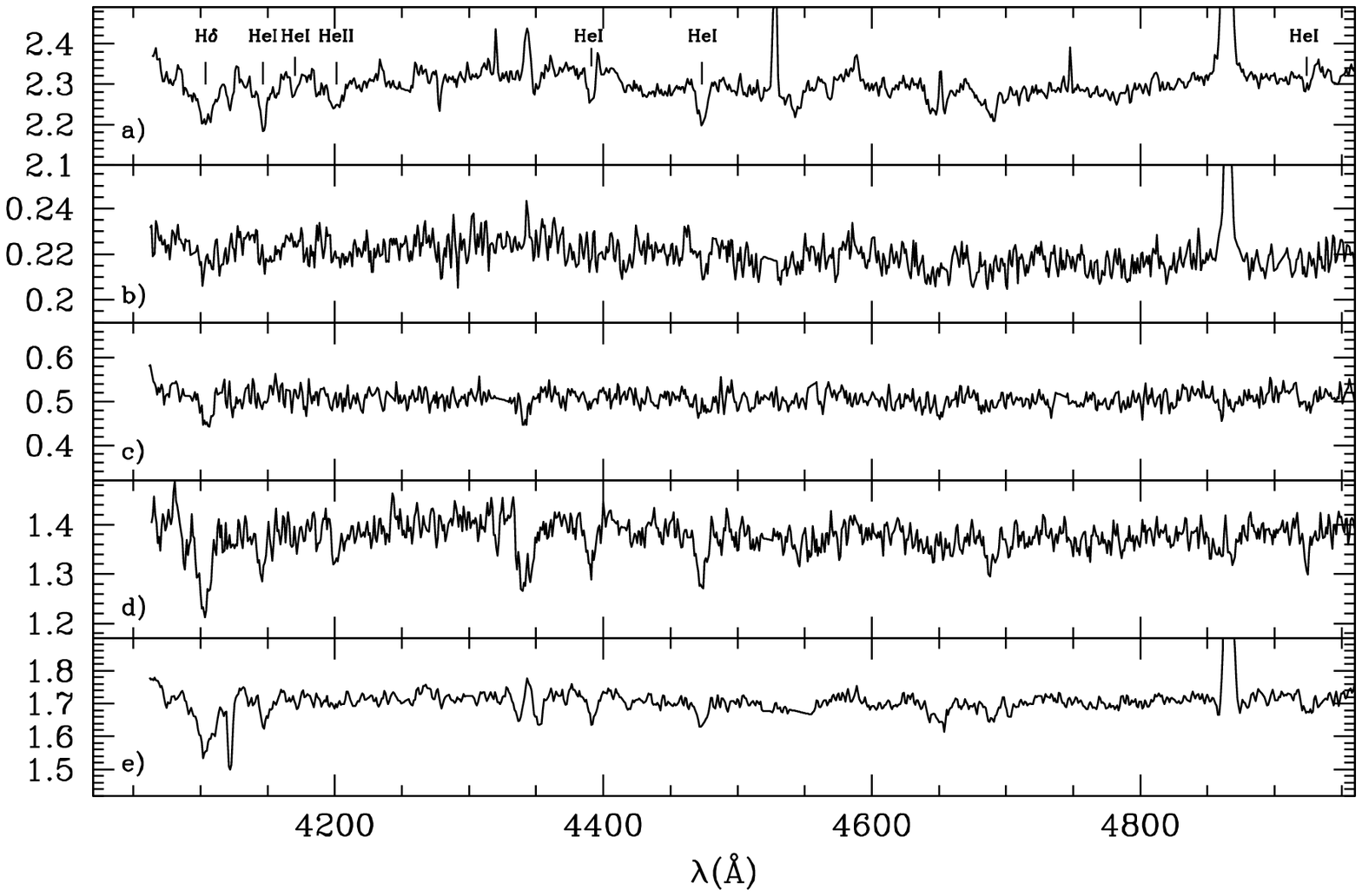}
\end{center}
\caption{ The top panel shows the RSS PG0900 grating flux-normalized spectra with arbitrary constants added to the flux to make the comparison easier. We plotted the whole spectral range and marked the main emission lines.
 a) XMM-J010147-715550 on 2016-09-07.
 The value of the normalization factor is $2.12\times 10^{-15}$ erg cm$^{-2}$ s$^{-1}$ \AA$^{-1}$.
b) XMM-J010147-7155 on 2016-09-29. The value of the
normalization factor is $2.05\times 10^{-16}$ erg cm$^{-2}$ s$^{-1}$ \AA$^{-1}$.
c) Suzaku-J0105-72 on 2016-09-20. The value of the
normalization factor is $4.38\times 10^{-16}$ erg cm$^{-2}$ s$^{-1}$ \AA$^{-1}$.
d) Suzaku-J0105-72 on 2016-10-25. The value of the
normalization factor is $1.23\times 10^{-15}$ erg cm$^{-2}$ s$^{-1}$ \AA$^{-1}$.
e) XMMU-J052016-692505 on 2016-10-25. The value of the normalization factor is $1.55\times 10^{-15}$ erg cm$^{-2}$ s$^{-1}$ \AA$^{-1}$.
The bottom panel zooms into the blue part of the spectrum ($\lambda < 4960$ \AA) to highlight the absorption lines.  The spectra were divided by the normalized continua and the fluxes are in unit of $10^{-15}$ erg cm$^{-2}$ s$^{-1}$ \AA$^{-1}$.  The most prominent absorption lines are marked. The letters correspond to the labels in the top panel.}
\label{fig2}
\end{figure}

\begin{figure}
\includegraphics[width=\columnwidth]{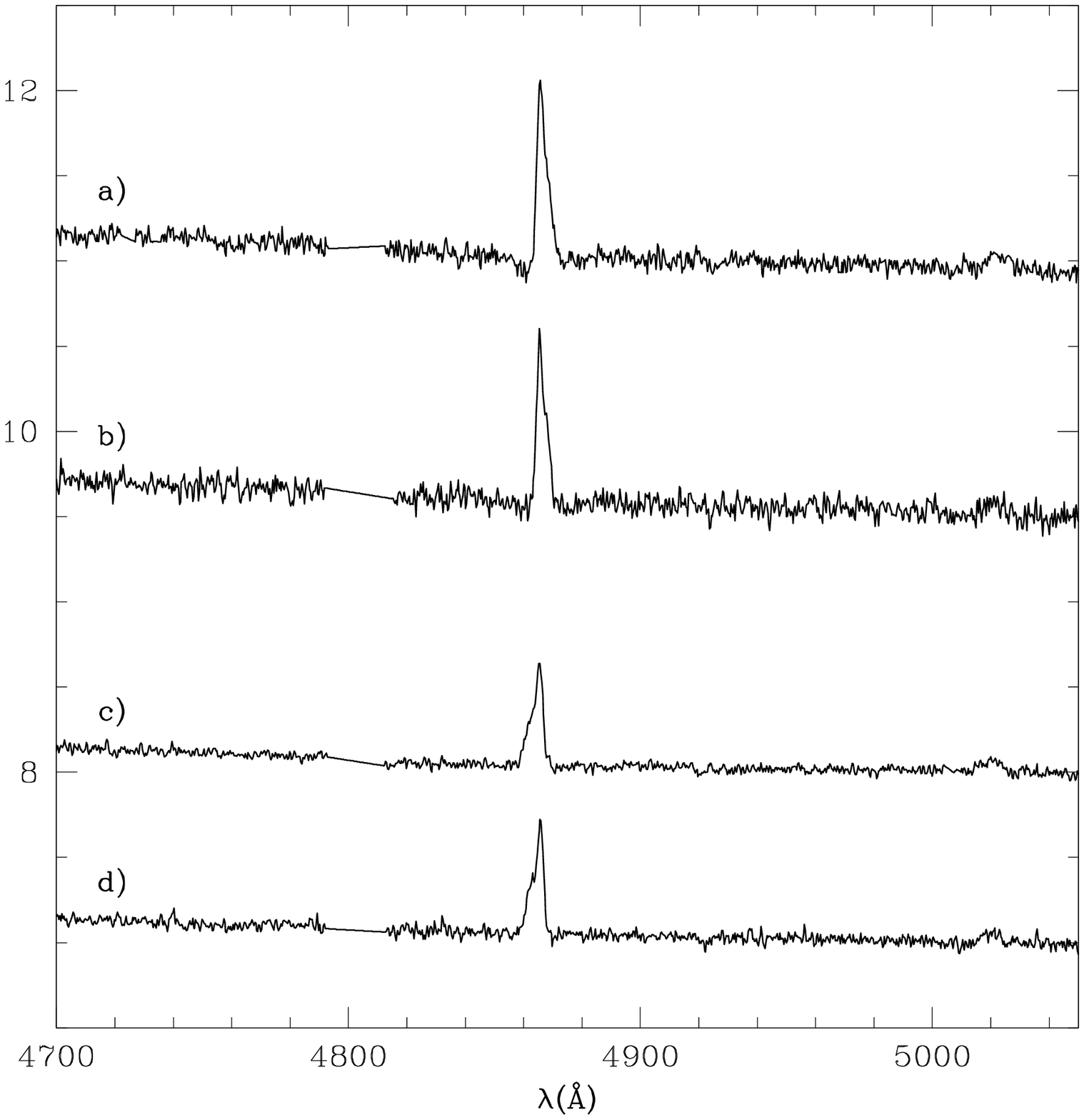}
\caption{Flux-normalized PG2300 grating spectra of XMMU-J052016-692505 (top) and MAXI-J0158-744 (bottom). The spectra were normalized using the flux at $\lambda = 4950$ \AA. An arbitrary constant was added to the flux in order to make the comparison easier. The spectral range is centered on H$\beta$ emission line, but it is also visible \ion{Fe}{2}\,$\lambda5018$ emission line. a) XMMU-J052016-692505 on 2016-09-20. The value of the normalization factor is $7.48\times10^{-16}$ erg cm$^{-2}$ s$^{-1}$ \AA$^{-1}$. b) XMMU-J052016-692505 on 2016-10-20. The value of the
normalization factor is $1.04\times10^{-15}$ erg cm$^{-2}$ s$^{-1}$ \AA$^{-1}$. c) MAXI-J0158-744 on 2016-10-10. The value of the
normalization factor is $1.37\times10^{-15}$ erg cm$^{-2}$ s$^{-1}$ \AA$^{-1}$. d) MAXI-J0158-744 on 2016-10-23. The value of the normalization factor is $1.61\times10^{-15}$ erg cm$^{-2}$ s$^{-1}$ \AA$^{-1}$. }
\label{fig3}
\end{figure}

All our RSS spectra show a steep blue continuum with Balmer hydrogen emission lines (H$\alpha$, H$\beta$ and H$\gamma$) as expected for Be stars; in some cases we detected  also \ion{He}{1} ($\lambda=5876$, 6678 and 7065 \AA) and \ion{Fe}{2} ($\lambda=4253$, 5018 and 5317 \AA) emission lines, and some hydrogen emission lines appear to arise from  shallower, broader absorption features. In some of the spectra, also the following hydrogen and helium absorption lines are present: H$\delta$, \ion{He}{1} ($\lambda=4143$, $4168$, $4387$, $4471$ and $4921$ \AA), \ion{He}{2} ($\lambda=4200$ \AA)).  All the RSS flux-normalized spectra are plotted in Fig.~\ref{fig2} (PG0900 grating) and Fig.~\ref{fig3} (PG2300 grating) to show which emission and absorption lines are present and measurable in which spectra. In both cases an arbitrary constant was added to the flux values to make the comparison easier. The values of the normalization flux are indicated in the captions of the figures. We listed in Table~\ref{tab3} the measured EWs for the emission lines of each spectrum.

Two spectra of XMMU-J010147.5-715550 obtained  22 days apart, clearly confirm the Be classification. In Fig~\ref{fig2} it is evident that the continuum slope did not change in the span of 3 weeks. 
Fewer emission lines were detected in the second spectrum, obtained with cloudy sky, thus noisier than the first one.  The stellar absorption lines, if present, are embedded in the noise.
 The S/N value for the continuum at $\lambda=5500$ \AA\ is 180 and 45 in each spectrum, respectively.  The EWs measured in both RSS spectra have approximately the same values,  however the EW of the hydrogen lines is smaller in HRS spectra obtained in 2017 than in the 2016 RSS spectra, while the EWs of the \ion{He}{1} lines show a marginal increase (see Table~\ref{tab3}).

 We found that the radial velocity is in the $150-190$ km s$^{-1}$ range, which is consistent with SMC membership \citep[systemic velocity varying from $88$ to $215$ km s$^{-1}$ according to][]{Stani1999}. In the RSS spectra we measured larger values of $\Delta v$ than in the HRS spectra.  If we compare  the $\sigma$ and $\Delta v$ values in Table~\ref{tab4} with EW values in Table~\ref{tab3}, we infer a slight increase of disk velocity from both H$\beta$ and H$\alpha$ between 2016 and 2017 corresponding to a decrease in EW values, as expected \citep{1988A&A...189..147H,Reig2016}, while there are no significant variations within timescales of weeks.
%%%%
%%%%
%
%%%%
\begin{table*}
\caption{The measured  equivalent widths: mean and standard deviation values are in \AA.}
{\tiny
\begin{tabular}{cccccccccccc}
\hline
Object & date &H$\gamma$ & H$\beta$ & H$\alpha$ & \ion{He}{1} & \ion{Fe}{2} & \ion{Fe}{2} & \ion{Fe}{2} & \ion{He}{1} & \ion{He}{1} & Notes\\
& &  & & & ($\lambda5876$) & ($\lambda4523$) & ($\lambda5018$) & ($\lambda5317$) & ($\lambda6678$) & ($\lambda7065$) &\\
\hline
XMMU-J010147.5-715550 & 2016-09-07 & 0.29$\pm$0.01 & 2.45$\pm$0.09 & 21.60$\pm$0.72 & 1.07$\pm$0.35 & 0.48$\pm$0.01  & 0.47$\pm$0.03 & 0.24$\pm$0.04 & 0.21$\pm$0.01 & 0.71$\pm$0.03 &\\
                      & 2016-09-29 & -- & 2.75$\pm$0.10 & 21.98$\pm$0.74 & 0.85$\pm$0.03 & -- & -- & 0.25$\pm$0.02 & -- & -- & (1)\\
                      & 2017-06-17 & -- & 2.03$\pm$0.20 & 18.20$\pm$1.77 & 1.20$\pm$0.33 & -- & -- & -- & -- & 0.69$\pm$0.15 & \\
                      & 2017-07-30 & -- & 2.24$\pm$0.19 & 16.22$\pm$1.55 & 1.62$\pm$0.28 & -- & -- & -- & -- & 1.21$\pm$0.12 & \\
                      & 2017-10-26 & -- & 2.25$\pm$0.12 & 16.74$\pm$1.01 & -- & -- & -- & -- & -- & 0.99$\pm$0.21 & \\
SUZAKU-J0105-72       & 2016-09-20 & -- & -- &3.23$\pm$0.15 & -- & -- & -- & -- & -- & -- & (2) \\
                      & 2016-10-19 & -- & -- &4.07$\pm$0.14 & -- & -- & -- & -- & -- & 0.54$\pm$0.06 & (2)\\
                      & 2017-07-14 & -- & 0.56$\pm$0.13 &2.97$\pm$0.46 & -- & -- & -- & -- & -- & -- &  \\
                      & 2017-08-26 & -- & 0.68$\pm$0.14 &3.12$\pm$0.42 & -- & -- & -- & -- & -- & -- &  \\
                      & 2017-10-26 & -- & 0.66$\pm$0.13 &3.12$\pm$0.30 & -- & -- & -- & -- & -- & -- &  \\
XMMU-J052016-692505   & 2016-09-20 & -- & 4.63$\pm$0.22 & -- & -- & -- & 0.89$\pm$0.04 & -- & -- & -- & (3)\\
                      & 2016-10-20 & -- & 3.47$\pm$0.08 & -- & -- & -- & 0.63$\pm$0.06 & -- & -- & -- & (3)\\
                      & 2016-10-25 & -- & 2.77$\pm$0.13 & 26.98$\pm$0.27 & 1.96$\pm$0.08 & -- & 0.53$\pm$0.08 & -- & 0.95$\pm$0.04 & 1.63$\pm$0.09 \\
                      & 2017-09-17 & -- & 2.84$\pm$0.28 & 19.65$\pm1.41$ & 5.41$\pm$0.72  & -- & -- & -- & -- & -- & \\
                      & 2017-10-22 & -- & 3.60$\pm$0.41 & 23.20$\pm2.32$ &   & -- & -- & -- & -- & 3.98$\pm$0.49 & \\
                      & 2017-10-29 & -- & 4.55$\pm$0.48 & 22.05$\pm1.35$ &   & -- & -- & -- & -- & 4.12$\pm$0.84 & \\
MAXI-J0158-744        & 2016-10-10 & -- & 2.87$\pm$0.09 & -- & -- & -- & 0.67$\pm$0.04 & -- & -- & -- & (3)\\
                      & 2016-10-23 & -- & 2.81$\pm$0.11 & -- & -- & -- & 0.55$\pm$0.03 & -- & -- & -- & (3)\\
                      & 2017-08-13 & -- & 3.68$\pm$0.24 & 26.07$\pm$2.35 & -- & -- & -- & -- & -- & 1.51$\pm$0.29 &\\
                      & 2017-09-17 & -- & 3.70$\pm$0.39 & 26.19$\pm$2.79 & -- & -- & -- & -- & -- & 1.11$\pm$0.25 &\\
                      & 2017-10-26 & -- & 3.72$\pm$0.42 & 29.77$\pm$1.83 & -- & -- & -- & -- & -- & 1.40$\pm$0.18 &\\
\hline
\label{tab3}
\end{tabular}}
\vspace{1ex}

\raggedright {\it Note} (1) This spectrum is noisier than the previous one and fewer emission lines are detectable; (2) These spectra are quite noisy, so only H$\alpha$ and \ion{He}{1} $\lambda7065$ are measurable; (3) These are higher resolution spectra: the spectral range is small and centered on H$\beta$.
\end{table*}%

\begin{table*}
\begin{center}
\caption{Kinematics and V/R.}
{\tiny
\begin{tabular}{cccccccc}
\hline
Object & Date&  Instrument & Radial velocity &  $\Delta v$ (H$\beta)^{1}$  & $\Delta v$ (H$\alpha)^{1}$& V/R (H$\beta$) & V/R (H$\alpha$)\\
       &     &       &    (km s$^{-1}$) & (km s$^{-1}$) & (km s$^{-1}$) & & \\
\hline
XMMU-J010147.5-715550  & 2016-09-07 & PG0900 & $186\pm22$  & $\mathbf{142\pm8}$ &  $\mathbf{153\pm7}$ &&\\
                       & 2016-09-29 & PG0900 & $208\pm114$ & $\mathbf{128\pm10}$ & $\mathbf{139\pm10}$ &&\\
                       & 2017-06-17 & HRS   & $149.1\pm0.4$ & $189\pm2$ & $\mathbf{171\pm6}$&$1.06\pm0.01$ &\\
                       & 2017-07-30 & HRS   & $153\pm4$ & $181\pm5$ & $\mathbf{160\pm5}$&$1.48\pm0.04$ &\\
                       & 2017-10-26 & HRS   & $177\pm1$ & $190\pm2$ & $\mathbf{170\pm3}$&$1.18\pm0.02$ &\\
SUZAKU-J0105-72  & 2016-09-20 & PG0900 & $196\pm75$ & & $\mathbf{100\pm11}$ &&\\
                 & 2016-10-19 & PG0900 & $170\pm86$ & & $\mathbf{127\pm17}$ &&\\
                 & 2017-07-14 & HRS   & $171\pm1 $ & $168\pm7$  & $\mathbf{115\pm10}$ &$1.40\pm0.07$ &\\
                 & 2017-08-26 & HRS   & $160\pm1$ $171\pm1$ & $164\pm4$  & $174\pm3$  &$0.71\pm0.02$  &$0.79\pm0.02$\\
                 & 2017-10-26 & HRS   & $188\pm3$ $177\pm1$ & $142\pm3$  & $198\pm3$  &$0.48\pm0.03$ &$0.64\pm0.01$\\
XMMU-J052016-692505   & 2016-10-25 & PG0900 & $282\pm54$ & $\mathbf{104\pm8}$ &$\mathbf{98\pm5}$ &&\\
                      & 2016-09-20 & PG2300 & $303\pm46$  & $165\pm11$ & &$1.96\pm0.23$ &\\
                      & 2016-10-20 & PG2300 & $287\pm109$ & $168\pm7$  & &$2.11\pm0.03$ &\\
                      & 2017-09-17 & HRS & $295\pm2$ & $135\pm4$ & $\mathbf{126\pm3}$ &$1.04\pm0.06$&\\
                      & 2017-10-22 & HRS & $285\pm1$ & $129\pm3$ & $\mathbf{116\pm4}$ &$0.80\pm0.08$&\\
                      & 2017-10-29 & HRS & $287\pm2$ & $125\pm4$ & $\mathbf{116\pm3}$ &$0.83\pm0.07$&\\
MAXI-J0158-744 & 2016-10-10 & PG2300 & $190\pm71$ & $209\pm5$  & &$0.46\pm0.01$ &\\
               & 2016-10-23 & PG2300 & $205\pm96$ & $180\pm14$ & &$0.57\pm0.07$ &\\
               & 2017-08-13 & HRS    & $167\pm2$ & $192\pm4$  & $\mathbf{173\pm5}$& $0.62\pm0.13$ &\\
               & 2017-09-17 & HRS    & $155\pm4$ & $200\pm7$  & $\mathbf{174\pm4}$& $0.59\pm0.02$ &\\
               & 2017-10-26 & HRS    & $143\pm1$ & $220\pm4$  & $\mathbf{175\pm5}$& $0.57\pm0.01$ &\\
\hline
\label{tab4}
\end{tabular}}
\end{center}
\vspace{1ex}

\raggedright {\it Note} {$^1 \Delta v$ is the velocity separation of
the red and blue peaks when the peaks are clearly visible,
 or it is assumed to be equivalent to the standard deviation of the Gaussian function ($\sigma$, values in boldface) in case of single-peaked profile.}
\end{table*}

Also for SUZAKU J0105-72 we were able to confirm the presence of a Be star, although the spectra did not allow detection of many  emission lines. We detected Be-typical Balmer hydrogen and helium emission and absorption lines on two different dates with a 29 days span in between.
The weather conditions were cloudy during the first run, so the resulting  spectrum is weaker and noisier than the second one (see Fig.~\ref{fig2}, with S/N of about 25 and 60, respectively, measured at $\lambda=5500$\AA). 
The two low-resolution spectra show similar continuum slopes, and  the small variations in the red and blue portions of the spectral range (of less than 10\%) are due to small differences in the sensitivity curves used for the photometric calibration.
In both spectra H$\beta$ is not clearly detectable, because it is stronger in absorption than in emission. The absorption should affect also the H$\alpha$ emission line, therefore our EW measurements for this object are lower limits.  The H$\alpha$ EW in the first low resolution spectrum is lower than in the second, while high resolution spectra show similar values .
 We measured radial velocity in the range $160-190$ km s$^{-1}$ in the RSS and HRS observations, which is consistent with 186 km s$^{-1}$ measured in a catalog by \citet{Evans2008}.
 We observed a small variation of $\Delta v$ in 2017 between July/August and October for H$\beta$. In 2016 the $\sigma$ values of H$\alpha$ do not seem to have varied. In 2017 we found variations between August and October while the EW values of H$\alpha$ and H$\beta$ are approximately constant. Therefore, for this object the expected relation between EW and $\Delta v$ does not seem to be valid.
Although small variations of EW are expected in case of small variations of $\Delta v$, according to the \citet{1988A&A...189..147H} relation, we do not have sufficient measurements to draw a definite conclusion. \citet{Reig2016} used data obtained over 15 years in their Fig. 7, so we suggest that much longer and possibly denser monitoring is necessary to draw conclusions. 

XMMU-J052016-692505 was observed only once with the RSS  and the PG0900 grating. Our spectrum is of better quality than the one of \citet[][]{Kahabka2006} and in addition to the emission lines of hydrogen, we detected \ion{He}{1} and \ion{Fe}{2} emission lines, making the Be star identification more solid.
 We found that also in this spectrum, the hydrogen emission lines are affected by absorption, so strong that H$\delta$ and H$\gamma$ are not measurable, therefore our measurements represent only lower limits for the EW of H$\alpha$ and H$\beta$.
The source was also observed twice  with a span of one month in between with the PG 2300 grating. 

The H$\beta$ profiles are similar and show a red asymmetry. We fitted them with two Gaussian functions.
In both spectra the H$\beta$ emission line is embedded in the corresponding absorption line.
 The radial velocity ($v \sim 280 - 300$ km s$^{-1}$) is consistent with LMC membership. 
For H$\beta$ we detected a small decrease of $\Delta v$ values between 2016 and 2017, and a small increase for the H$\alpha$ line. EW and $\Delta v$ of H$\alpha$ seem to follow the \citet{1988A&A...189..147H} relation, even if we have only four data points. For H$\beta$ we have more data points but the correlation is less definite, even if there is an indication of decreasing EW for higher values of $\Delta v$.

Because the published spectrum of  MAXI-J0158-744 \citep[][]{Li2012} is of good quality, we focused on improving the resolution on the spectral range of H$\beta$ and of the \ion{He}{2} line at 4686 \AA, using only the PG2300 grating. It was observed twice  with a 13 day span in between (Fig.~\ref{fig3}).
The spectral range is centered on the H$\beta$ emission line, which is asymmetric with a blue wing in both spectra. The \ion{Fe}{2} $\lambda 5018$ is detected in our RSS spectra, but  it was not measured by \citet[][]{Li2012}. It is noisy and asymmetric in both our spectra. The weak \ion{He}{2} emission line at 4686 \AA, detected by \citet[][]{Li2012} is instead not observed in our spectra.

The range of radial velocity is between 140 and 200 km s$^{-1}$. The H$\beta$ profiles do not vary much, and we fitted them with two Gaussian functions.  $\Delta v$, for both H$\alpha$ and H$\beta$ lines, does not show variations. The EW  do not vary for  both H$\beta$ and the \ion{Fe}{2} emission lines in the RSS spectra, but  larger values were measured for H$\beta$ in the HRS ones. The H$\alpha$ and \ion{He}{1} EW values are consistent within the errors. EW and $\Delta v$ of both H$\alpha$ and H$\beta$ do not follow the expected relation, but also we do not have enough data to draw a definite conclusion.

\subsection{Analysis of the emission line profiles}

\begin{figure}[htb!]
\includegraphics[width=\columnwidth]{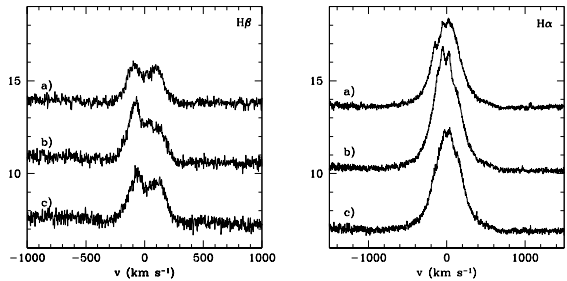}
\caption{The Balmer
emission line profiles of XMMU-J010147.5-715550. The fluxes are in unit of $10^{-15}$ erg cm$^{-2}$ s$^{1}$ \AA$^{-1}$.
Arbitrary constants were added to the flux to make the comparison easier.
 Left panel: H$\beta$ emission line profile.
a) Spectrum on 2017-06-17;
b) spectrum on 2017-07-30;
c) spectrum on 2017-10-26.
Right panel: H$\alpha$ emission line profile. The letters refer to the same dates.}
\label{fig4}
\end{figure}

\begin{figure}[htb!]
\includegraphics[width=\columnwidth]{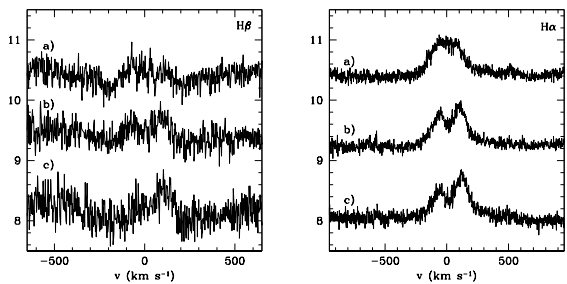}
\caption{Variability of the hydrogen emission line profiles of SUZAKU J0105-72. The fluxes are in unit of $10^{-15}$ erg cm$^{-2}$ s$^{1}$ \AA$^{-1}$. Arbitrary constants were added to the flux to make the comparison easier. Left panel: H$\beta$ emission line profile.
  a) Spectrum on 2017-07-14;
  b) spectrum on 2017-08-26;
  c) spectrum on 2017-10-26.
  Right panel: H$\alpha$ emission line profile. The letters refer to the same dates.} 
\label{fig5}
\end{figure}

\begin{figure}[htb!]
\includegraphics[width=\columnwidth]{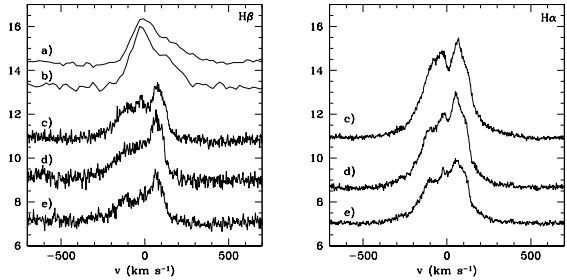}
\caption{The hydrogen emission line profiles of XMMU-J052016-692505. The fluxes are in unit of $10^{-15}$ erg cm$^{-2}$ s$^{1}$ \AA$^{-1}$. Arbitrary constants were added to the flux to make the comparison easier. a) and b) H$\beta$ fluxes, obtained with PG2300, were multiplied by a factor of 2.5 to make them comparable to those obtained with the
  HRS. Left panel: H$\beta$ emission line profile.
  a) PG2300 spectrum on 2016-09-20;
  b) PG2300 spectrum on 2016-10-20;
  c) HRS spectrum on 2017-09-17;
  d) HRS spectrum  on 2017-10-22;
  e) HRS spectrum on 2017-19-29.
  Right panel: H$\alpha$ emission line profile. The letters refer to the same dates.}
\label{fig6}
\end{figure}

\begin{figure}[htb!]
\includegraphics[width=\columnwidth]{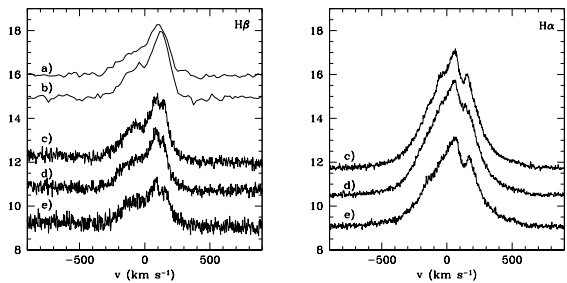}
\caption{Variability of the hydrogen emission line profiles of MAXI-J0158-744. The fluxes are in unit of $10^{-15}$ erg cm$^{-2}$ s$^{1}$ \AA$^{-1}$. Arbitrary constants were added to the flux to make the comparison easier. a) and b) H$\beta$ fluxes, obtained with PG2300, were multiplied by a factor of 2.5 to make them comparable to those obtained with HRS. Left panel: H$\beta$ emission line profile.
  a) PG2300 spectrum on 2016-10-10;
  b) PG2300 spectrum on 2016-10-23;
  c) HRS spectrum on 2017-08-13;
  d) HRS spectrum on 2017-09-17;
  e) HRS spectrum on 2017-10-26.
  Right panel: H$\alpha$ emission line profile. The letters refer to the same dates.}
\label{fig7}
\end{figure}

The higher resolution (PG2300 and HRS) spectra allow us to study the emission line profiles, and to compare them at different epochs.
We plotted the profiles in velocity space in Figs.~\ref{fig4}-\ref{fig7}, assuming as systemic velocity of the SMC sources the average radial velocity of the SMC, 158 km s$^{-1}$, and we adopted 278 km s$^{-1}$ for XMMU-J052016.0-692505, which is in the LMC.

 XMMU-J010147.5-715550 was observed with the HRS at three different epochs.
The H$\beta$ line shows two peaks with V/R $\sim 1$ in the first epoch spectrum,
 then the ratio increased and decreased again.
The H$\alpha$ line does not show a distinct double-peaked profile and it is quite
 different from the H$\beta$. It is asymmetric, with three peaks in the first epoch, but 
with a ``wine-bottle'' profile with two peaks in the second and third epoch (see Fig.~\ref{fig4}).

%%%
%%%

The SUZAKU J0105-72 HRS spectra are noisier than those of the other sources,
 but show weak hydrogen emission lines than in the spectra of other sources. Nonetheless, both H$\beta$ and H$\alpha$ have a clear double peaked profile, which is almost flat at the first epoch, but still allows a double Gaussian fitting, 
at least for H$\beta$. The V/R ratio for H$\beta$ 
 decreased between 2017 July and two observation in 2017 October,  and
 in 2017 October we could measure it also for H$\alpha$,
 finding agreement with the H$\beta$ values. 
We notice that the profiles of both lines underwent
 a similar evolution through the three epochs (see Fig.~\ref{fig5}).

%%%
%%%

For XMMU-J052016-692505 we plotted the
H$\beta$ line observed with PG2300 in 2016, and 
 both H$\beta$ and H$\alpha$ for the HRS 2017 observations.
This is the most interesting object of our sample,
 because the emission lines profiles show a clear evolution (see Fig.~\ref{fig6}).
 In  2016 H$\beta$ had an asymmetric profile with a red wing,
 possibly due to a less bright, unresolved R peak. 
In 2017, the profile had changed shape with a red peak and a blue wing,
which is in fact a blue peak of diminished
 intensity, 
rather stable for the rest of 2017. The V/R values decreased between 2017 July and
 the two 2017 October observations, and
  also the H$\alpha$ shows evolution in 2017. 
 While two peaks may be present,
 but are difficult to resolve in the first epoch, a triple peaked profile appears in the second and third epoch. The general shape of H$\alpha$ resembles that of H$\beta$,
with a less intense blue peak, and it is possible that we did not
 resolve three peaks for the H$\beta$ line.

%%%
%%%

Also for MAXI-J0158-744 we took both PG2300 and HRS spectra for H$\beta$. There is little evolution of the line profiles, which show a red peak permanently brighter than the blue one (see Fig.~\ref{fig7}). Also in this system, the H$\alpha$ lines profiles are different from the H$\beta$ profiles. The profiles are all asymmetric with a clear excess in the blue side. Three peaks can be distinguished in the first H$\alpha$ spectrum, although only two are prominent, and the same may be true in the last H$\beta$ spectrum.

%%%
%%%

\section{Discussion}\label{sec:discussion}

 We have confirmed the Be classification of all four optical counterparts of these
 SSS, whose spectra have the typically double peaked emission
 lines of Be stars \citep[see, among others,][]{Struve1945, Dachs+92,
Hanuschik1996, 2018AJ....155...53L} and radial
 velocities that are consistent with Magellanic Clouds' membership. These four objects
 are clearly similar to the ``emission-line Be stars''
 classified by \citet{Porter2003} rather than to ``shell Be stars'', because the
 central reversal point between the peaks of the Balmer hydrogen lines and some other lines
 of \ion{He}{1} and \ion{Fe}{2} is always above the flux level of the stellar photosphere
 (however, the shell Be may only be an effect of inclination, see \citealt{Hanuschik1996}).
It is interesting that the excretion disk seems persistent over the years; in
 many Be stars instead the emission line spectrum is not observed at all epochs.
Thus, the hypothesis of \citet{Sturm2012} of a transient excretion disk for 
XMMU-J010147.5-715550 seems unfounded.
We have not detected any optical emission line indicating high ionization or excitation potential; most notably the \ion{He}{2} $\lambda$4686 line is missing, showing that, most likely, there is no accretion disk. Not many clear-cut cases of accretion disks in Be binaries are known, but one is a rare Be+black hole system, MWC 656, and in that system the  \ion{He}{2} $\lambda$ 4686 line is clearly detected  \citep{Casares2012, Casares2014}. 
 
 Another candidiate Be+black hole system is $\phi$ Per \citep{Poeckert1979}, but this is not even a detectable  X-ray source. However, the perfect spatial
 coincidence of five Be stars with SSS cannot be coincidental,
 so these Be stars should have a binary compact object companion from which
 the SSS emission originates. We suggest that the
 optical spectra imply that these binaries are similar
 to the many neutron star
 X-ray binaries in which evidence
 of an accretion disk is missing: the neutron star accretes either when it crosses the
 excretion disk of the Be star, or when there is renewed activity of the Be that
 increases the disk size and amount of material. In several 
 known high mass X-ray binaries (HMXB) 
 there is evidence that a temporary accretion disk is formed that does
 not remain for long \citep{Reig2011} and although we cannot rule out that the
 accretion disk is ever formed for the systems studied here,
 it seems very unlikely that a disk feeding a black hole in a 
 super-Eddington accretion regime  can have been 
 formed and disappeared in such short time that it was never
 observed for our sources.
 There are some interesting observable properties
 in these objects, that should be
 monitored in the future. We discuss here the points that are
 relevant for future programs. 

 1) In absence of an accretion disk, the velocity difference $\Delta v$ between the red and blue peak of the most prominent emission lines, that we obtained from the H$\beta$ line, and in one case also for the H$\alpha$ line, is not indicative of orbital motion;  it indicates the excretion disk rotation velocity, as commonly assumed for Be stars. 
 Although we measured variations of the line profiles and the V/R ratio within weeks, $\Delta v$  varied within few weeks only for SUZAKU J0105-72.
In the other objects, velocity variations occurred on a one year timescale. 
 The derived velocity is quite lower ($\Delta v = 100 - 200$ and $\Delta v = 100 - 220$ km s$^{-1}$ for H$\alpha$ and H$\beta$, respectively, corresponding to $v_d \sin i$ of about 50-100 and 50-110 km s$^{-1}$) than the average rotation velocity of galactic  Be stars of different types \citep[223 km s$^{-1}$ for early Be type with weak emission lines, 270 km s$^{-1}$ for all the others, see][]{Briot1986}. 
\citet{Hanuschik1996} found that in many Be the disk is differentially rotating, with its axis aligned with the stellar rotation axis and a velocity law which is radially decreasing. Indeed, we know that the disks of both classic and binary Be are in Keplerian rotation \citep{2012A&A...538A.110M, 2013A&ARv..21...69R}, but the latter are denser than the former and truncated \citep{Reig2016}. In a disk in Keplerian rotation, the radius is proportional to the inverse square of the velocity.  
If R$_*$ is the radius of the Be star and v$_*$ is its rotation velocity,  the disk radius $r_{\rm d}$ can be inferred from $\Delta v$ as $$(r_{\rm d}/R_*)=(2 v_* \sin i / \Delta v)^{1/j}$$ where of course 1/j=2 for Keplerian rotation \citep{Reig2016}.
By applying this formula and using our mean value $\Delta v = 150$ km s$^{-1}$ and the mean value of $v_* \sin i$ from \citet{Briot1986} (250 km s$^{-1}$), we obtain r$_d \sim 11$ R$_*$, implying a relatively small size disk, compared to the values for classic Be stars \citep[$14-22$ R$_*$,][]{Reig2016}.

2) We found that the emission lines' profile change in time, but while for  SUZAKU J0105-72 H$\beta$ and H$\alpha$ have similar profiles in the same spectrum, in
 two other sources, XMMU-J010147.5-715550 and MAXI-J0158-744,
 each of these two emission lines is different from the other at a given epoch.
 In MAXI-J0158-744, there are only small differences in the line profiles at a given
 epoch. We note that in XMMU-J052016-692505 there is 
 even an inversion in the height of the blue and red
 peak in H$\beta$. Because H$\alpha$ has a lower excitation potential,
 it is supposed to arise in a cooler region than H$\beta$. According to \citet{Sigut2007}, the thermal structure of the Be disk involves higher temperature in the "edge" region at high azimuthal height. Moreover, the optical depth of H$\alpha$ is higher than that of H$\beta$, so the H$\alpha$ line appears stronger (forming over a larger radial portion of the disk).

 In a recent article \citet{Panoglou2018} calculate models of excretion
 disks of binary Be stars with radiative transfer, and find that binary tidal effects
 cause phase locked variability of the V/R ratio of the Balmer lines, with two maxima per
 orbital cycle. Thus, we may be able to infer the orbital period of our systems by
 measuring the V/R with a short cadence, of one or two weeks perhaps. 

3) \citet{Reig2011} and \citet{Reig2016} found interesting  correlations
 of the EW of the H$\alpha$ line, in all Be stars with excretion disk size, 
 and in HMXBs with the orbital period. \citet{Reig2011} 
proposed that the orbital period
 determines the size of the disk, which is truncated by the tidal interaction of the
 secondary. Since the H$\alpha$ line originates in the cooler, outer zone
 of the disk, it is easily conceivable 
 that its equivalent width is inversely proportional to
 the disk size.  Fig. 15 of \citet{Reig2011} shows that the maximum EW(H$\alpha$) in HMXB 
 is proportional to the orbital period.
 In our cases, for Suzaku J0105-72 the orbital period should be of the order of less than 10 days (EW$\sim 4$ \AA), while it should be 2-4 months for our other 3 sources 
(EW in the range 22--30 \AA). These are actually lower limits, in fact the model assumes that the disk is dense enough to reach the tidal radius of the binary. This may not be true if there are periods of interrupted, or decreased  mass outflow from the Be star. In any case, even having upper limits on the period is useful as a benchmark to plan observations with a suitable cadence and measure the orbital periods of these binaries.
However, in short orbital period systems the EW may
 be variable. \citet{Reig2016} studied
 also the variability of the H$\alpha$ EW in HMXB and  
 found that systems with short orbital periods are affected by the neutron star
 tidal truncation of the disk more often and more strongly, so the disk does not
 have a stable configuration and its density  changes quickly. 

 4) Finally, we should take into account also  two 
 suggestions by \citet{Raguzova2001}. The first concerns the 
\ion{He}{1} emission lines. These lines require larger ionization potential than the Balmer hydrogen lines, are typical of Be with a WD companion because they are formed in the Str\"omgren sphere near the WD, and indicate the WD motion. \citet{Raguzova2001} calculated that, if the orbital period exceeds about 3 months, the \ion{He}{1} lines should have a semi-amplitude of the velocity curve due to orbital motion  of 70 km s$^{-1}$ that should be well measurable.
 Both this type of measurement, and that of the broadening of the Be absorption lines to determine the rotational velocity of the Be star, possibly require larger telescope collecting area, and certainly require longer exposures, a more frequent sampling and/or higher spectral resolution than those  we used in the present project. Raguzova also predicted
 that the semi-amplitude of the velocity curve  of the Balmer lines, associated with the excretion disk around
 the Be star, would only be 10-20 km s$^{-1}$.
 Of course in the Magellanic Clouds the measured radial velocity of the Balmer lines is  due to the systemic velocity in the region of the Clouds in which each star is located plus the component due to the orbital motion. To obtain the semi-amplitude from the velocity curve, higher spectral resolution and measurements repeated with more frequency (every week or once a day) are needed. 

\section{Conclusions}\label{sec:summary}

In the previous Section, we have outlined possible photometric and spectroscopic
 observations that may yield  measurements of the orbital period, of the WD rotation period, and finally would allow deriving the masses of the Be star and the compact object. While certainly there is more interesting work ahead before completely understanding the nature of these systems, we are not aware of a phenomenology that would make neutron stars
 intermittent or transient supersoft X-ray emitters
 without presenting also harder X-ray states and/or
 a much harder X-ray ``tail'' in addition to the SSS spectrum
\citep[see][]{Kylafis1993,
 Kylafis1996, Kohno2000}.  The SSS phenomenology
 seems rather typical of burning WDs or, more
 rarely, of black hole with extended accretion disks. However,
 also the Be+black hole scenario
is highly unlikely: large, luminous, steady and ``cool'' disks,
 like the one in the ultra-luminous
 SSS in M101 \citep{Liu2013} and  perhaps in the M81
 source \citep{Liu2015}, do seem
 to produce a supersoft X-ray spectrum, but they are
 persistent SSS and there is a prominent \ion{He}{2} line in the case of M101.  
 We suggest that the absence of this emission line and 
 the short duration of the SSS phases in our systems
 cannot be reconciled with an accreting black hole as compact object.

 The clear association of 5 Be counterparts out of about 30 SSS in the Magellanic Clouds suggests
 that such binaries may be common. The fact that many SSS in M31 and in galaxies
outside the Local Group  are in star forming regions \citep[][]{Orio2010, DiStefano2003}
also indicates that some
 transient or recurrent SSS may be the detectable manifestation of Be stars with WD
companions, thanks to the luminous X-ray property of shell hydrogen burning,
ignited perhaps intermittently as the X-ray observations of these four sources seem to indicate.
Be+WD star systems, though difficult to discover, may be common according to the works
we mentioned in the Introduction, so this may be a whole new class of interacting binary
progenitors of type Ia supernovae.

The model of massive WDs accreting from Be stars and burning hydrogen has been proposed
 to explain several cases of SSS, but the evolution of such systems has not been explored yet.
Would they eject the accreted material in nova type outburst? A nova in a Be star
systems may easily be missed, having an amplitude of less than 3 magnitudes,
 not even comparable with the optical novae we know in low mass systems. Moreover, if the
WD accretes and burns only sporadically during the orbital cycle, mass ejection may be
rare or impossible and the already massive WD would grow towards the Chandrasekhar mass.
There is one more parameter to take into account, and it is the WD temperature at the
 beginning of mass accretion. Because the WDs in these systems are very hot, they have
less degenerate
accreted envelopes and most likely, even if they undergo nova outbursts,
they eject less mass in nova events and do not undergo  phases of helium shell
flashes \citep{Hillman2016}, possibly continuing to accrete more material than they can ever
eject and being on a path towards SNe Ia explosion or accretion induced collapse into
 neutron stars.

 The dependence of the type Ia SN rate on star formation rate is known since the pioneering
 works of \citet[][]{Tammann1977, Tammann1978, Oemler1979} and has been confirmed and
explored time and again by different authors \citep[see][]{Mannucci2006, Sullivan2006};
it will be very interesting to explore how this new class of progenitors
 fits into galactic evolution and explains SNe Ia in young populations. In evaluating the
 statistics and the time for accretion towards the
Chandrasekhar mass we must remember that the Be+WD binaries are not long lived, but they
last longer than the numerous neutron star+Be star systems we know and their short-lived
 SSS phases make them difficult to identify even in X-rays. 

 Following the work of \citet{Panoglou2018}, the changing
 profile of the optical emission lines in Be stars
 is an indication of binarity. We also
 note that appearance of a  triple-peak profile, observed
 sporadically in our spectra, is not a proof against binarity, according to the above authors. In fact the Be binary zeta $\tau$ does present triple peaks \citep{Carciofi2009}. Perhaps the way to detect the presence of a WD in a Be star without a known orbital companion may be cyclic variability of V/R (due to the orbital period of the otherwise undetectable WD).
This would be  a more promising way of discovering Be+WD systems than ``waiting'' for a SSS flare.
If we want to build statistics of Be+WD binaries, the X-rays have probably only
 shown the tip of the iceberg, while optical studies are a  more feasible avenue to increase the number of known systems.

\acknowledgments
We are greatful to the anonymous referee for useful comments and suggestions that helped to improve the quality of the paper. All the observations reported in this paper were obtained with the Southern African Large Telescope (SALT).

\bibliographystyle{apj}
\bibliography{biblio}

\begin{thebibliography}{}
\expandafter\ifx\csname natexlab\endcsname\relax\def\natexlab#1{#1}\fi

\bibitem[{{Antoniou} \& {Zezas}(2016)}]{Antoniou2016}
{Antoniou}, V., \& {Zezas}, A. 2016, \mnras, 459, 528

\bibitem[{{Bramall} {et~al.}(2010){Bramall}, {Sharples}, {Tyas}, {Schmoll},
  {Clark}, {Luke}, {Looker}, {Dipper}, {Ryan}, {Buckley}, {Brink}, \&
  {Barnes}}]{2010SPIE.7735E..4FB}
{Bramall}, D.~G., {Sharples}, R., {Tyas}, L., {et~al.} 2010, in \procspie, Vol.
  7735, Ground-based and Airborne Instrumentation for Astronomy III, 77354F

\bibitem[{{Bramall} {et~al.}(2012){Bramall}, {Schmoll}, {Tyas}, {Clark},
  {Younger}, {Sharples}, {Dipper}, {Ryan}, {Buckley}, \&
  {Brink}}]{2012SPIE.8446E..0AB}
{Bramall}, D.~G., {Schmoll}, J., {Tyas}, L.~M.~G., {et~al.} 2012, in \procspie,
  Vol. 8446, Ground-based and Airborne Instrumentation for Astronomy IV, 84460A

\bibitem[{{Briot}(1986)}]{Briot1986}
{Briot}, D. 1986, \aap, 163, 67

\bibitem[{{Buckley} {et~al.}(2006){Buckley}, {Swart}, \&
  {Meiring}}]{2006SPIE.6267E..0ZB}
{Buckley}, D.~A.~H., {Swart}, G.~P., \& {Meiring}, J.~G. 2006, in \procspie,
  Vol. 6267, Society of Photo-Optical Instrumentation Engineers (SPIE)
  Conference Series, 62670Z

\bibitem[{{Burgh} {et~al.}(2003){Burgh}, {Nordsieck}, {Kobulnicky}, {Williams},
  {O'Donoghue}, {Smith}, \& {Percival}}]{2003SPIE.4841.1463B}
{Burgh}, E.~B., {Nordsieck}, K.~H., {Kobulnicky}, H.~A., {et~al.} 2003, in
  \procspie, Vol. 4841, Instrument Design and Performance for Optical/Infrared
  Ground-based Telescopes, ed. M.~{Iye} \& A.~F.~M. {Moorwood}, 1463--1471

\bibitem[{{Carciofi} {et~al.}(2009){Carciofi}, {Okazaki}, {Le Bouquin}, {{\v
  S}tefl}, {Rivinius}, {Baade}, {Bjorkman}, \& {Hummel}}]{Carciofi2009}
{Carciofi}, A.~C., {Okazaki}, A.~T., {Le Bouquin}, J.-B., {et~al.} 2009, \aap,
  504, 915

\bibitem[{{Casares} {et~al.}(2014){Casares}, {Negueruela}, {Rib{\'o}}, {Ribas},
  {Paredes}, {Herrero}, \& {Sim{\'o}n-D{\'{\i}}az}}]{Casares2014}
{Casares}, J., {Negueruela}, I., {Rib{\'o}}, M., {et~al.} 2014, \nat, 505, 378

\bibitem[{{Casares} {et~al.}(2012){Casares}, {Rib{\'o}}, {Ribas}, {Paredes},
  {Vilardell}, \& {Negueruela}}]{Casares2012}
{Casares}, J., {Rib{\'o}}, M., {Ribas}, I., {et~al.} 2012, \mnras, 421, 1103

\bibitem[{{Crause} {et~al.}(2014){Crause}, {Sharples}, {Bramall}, {Schmoll},
  {Clark}, {Younger}, {Tyas}, {Ryan}, {Brink}, {Strydom}, {Buckley},
  {Wilkinson}, {Crawford}, \& {Depagne}}]{2014SPIE.9147E..6TC}
{Crause}, L.~A., {Sharples}, R.~M., {Bramall}, D.~G., {et~al.} 2014, in
  \procspie, Vol. 9147, Ground-based and Airborne Instrumentation for Astronomy
  V, 91476T

\bibitem[{{Dachs} {et~al.}(1992){Dachs}, {Hummel}, \& {Hanuschik}}]{Dachs+92}
{Dachs}, J., {Hummel}, W., \& {Hanuschik}, R.~W. 1992, \aaps, 95, 437

\bibitem[{{Di Stefano} \& {Kong}(2003)}]{DiStefano2003}
{Di Stefano}, R., \& {Kong}, A.~K.~H. 2003, \apj, 592, 884

\bibitem[{{Di Stefano} \& {Kong}(2004)}]{DiStefano2004}
---. 2004, \apj, 609, 710

\bibitem[{{Evans} \& {Howarth}(2008)}]{Evans2008}
{Evans}, C.~J., \& {Howarth}, I.~D. 2008, \mnras, 386, 826

\bibitem[{{Evans} {et~al.}(2004){Evans}, {Howarth}, {Irwin}, {Burnley}, \&
  {Harries}}]{Evans2004}
{Evans}, C.~J., {Howarth}, I.~D., {Irwin}, M.~J., {Burnley}, A.~W., \&
  {Harries}, T.~J. 2004, \mnras, 353, 601

\bibitem[{{Gies} {et~al.}(2008){Gies}, {Dieterich}, {Richardson}, {Riedel},
  {B.~L.~Team}, {McAlister}, {Bagnuolo}, {Grundstrom}, {{\v S}tefl},
  {Rivinius}, \& {Baade}}]{Gies2008}
{Gies}, D.~R., {Dieterich}, S., {Richardson}, N.~D., {et~al.} 2008, \apjl, 682,
  L117

\bibitem[{{Greiner} {et~al.}(1991){Greiner}, {Hasinger}, \&
  {Kahabka}}]{Greiner1991}
{Greiner}, J., {Hasinger}, G., \& {Kahabka}, P. 1991, \aap, 246, L17

\bibitem[{{Greiner} {et~al.}(1996){Greiner}, {Schwarz}, {Hasinger}, \&
  {Orio}}]{Greiner1996}
{Greiner}, J., {Schwarz}, R., {Hasinger}, G., \& {Orio}, M. 1996, \aap, 312, 88

\bibitem[{{Hanuschik}(1996)}]{Hanuschik1996}
{Hanuschik}, R.~W. 1996, \aap, 308, 170

\bibitem[{{Hanuschik} {et~al.}(1988){Hanuschik}, {Kozok}, \&
  {Kaiser}}]{1988A&A...189..147H}
{Hanuschik}, R.~W., {Kozok}, J.~R., \& {Kaiser}, D. 1988, \aap, 189, 147

\bibitem[{{Henze} {et~al.}(2011){Henze}, {Pietsch}, {Haberl}, {Hernanz},
  {Sala}, {Hatzidimitriou}, {Della Valle}, {Rau}, {Hartmann}, \&
  {Burwitz}}]{Henze2011}
{Henze}, M., {Pietsch}, W., {Haberl}, F., {et~al.} 2011, \aap, 533, A52

\bibitem[{{Henze} {et~al.}(2014){Henze}, {Pietsch}, {Haberl}, {Della Valle},
  {Sala}, {Hatzidimitriou}, {Hofmann}, {Hernanz}, {Hartmann}, \&
  {Greiner}}]{Henze2014}
---. 2014, \aap, 563, A2

\bibitem[{{Hillman} {et~al.}(2016){Hillman}, {Prialnik}, {Kovetz}, \&
  {Shara}}]{Hillman2016}
{Hillman}, Y., {Prialnik}, D., {Kovetz}, A., \& {Shara}, M.~M. 2016, \apj, 819,
  168

\bibitem[{{Kahabka} {et~al.}(2006){Kahabka}, {Haberl}, {Payne}, \&
  {Filipovi{\'c}}}]{Kahabka2006}
{Kahabka}, P., {Haberl}, F., {Payne}, J.~L., \& {Filipovi{\'c}}, M.~D. 2006,
  \aap, 458, 285

\bibitem[{{Kobulnicky} {et~al.}(2003){Kobulnicky}, {Nordsieck}, {Burgh},
  {Smith}, {Percival}, {Williams}, \& {O'Donoghue}}]{2003SPIE.4841.1634K}
{Kobulnicky}, H.~A., {Nordsieck}, K.~H., {Burgh}, E.~B., {et~al.} 2003, in
  \procspie, Vol. 4841, Instrument Design and Performance for Optical/Infrared
  Ground-based Telescopes, ed. M.~{Iye} \& A.~F.~M. {Moorwood}, 1634--1644

\bibitem[{{Kohno} {et~al.}(2000){Kohno}, {Yokogawa}, \& {Koyama}}]{Kohno2000}
{Kohno}, M., {Yokogawa}, J., \& {Koyama}, K. 2000, \pasj, 52, 299

\bibitem[{{Kourniotis} {et~al.}(2014){Kourniotis}, {Bonanos}, {Soszy{\'n}ski},
  {Poleski}, {Krikelis}, {Udalski}, {Szyma{\'n}ski}, {Kubiak},
  {Pietrzy{\'n}ski}, {Wyrzykowski}, {Ulaczyk}, {Koz{\l}owski}, \&
  {Pietrukowicz}}]{Kourniotis2014}
{Kourniotis}, M., {Bonanos}, A.~Z., {Soszy{\'n}ski}, I., {et~al.} 2014, \aap,
  562, A125

\bibitem[{{Kylafis}(1996)}]{Kylafis1996}
{Kylafis}, N.~D. 1996, in Lecture Notes in Physics, Berlin Springer Verlag,
  Vol. 472, Supersoft X-Ray Sources, ed. J.~{Greiner}, 41

\bibitem[{{Kylafis} \& {Xilouris}(1993)}]{Kylafis1993}
{Kylafis}, N.~D., \& {Xilouris}, E.~M. 1993, \aap, 278, L43

\bibitem[{{Labadie-Bartz} {et~al.}(2018){Labadie-Bartz}, {Chojnowski},
  {Whelan}, {Pepper}, {McSwain}, {Borges Fernandes}, {Wisniewski},
  {Stringfellow}, {Carciofi}, {Siverd}, {Glazier}, {Anderson}, {Caravello},
  {Stassun}, {Lund}, {Stevens}, {Rodriguez}, {James}, \&
  {Kuhn}}]{2018AJ....155...53L}
{Labadie-Bartz}, J., {Chojnowski}, S.~D., {Whelan}, D.~G., {et~al.} 2018, \aj,
  155, 53

\bibitem[{{Lamb} {et~al.}(2016){Lamb}, {Oey}, {Segura-Cox}, {Graus}, {Kiminki},
  {Golden-Marx}, \& {Parker}}]{Lamb2016}
{Lamb}, J.~B., {Oey}, M.~S., {Segura-Cox}, D.~M., {et~al.} 2016, \apj, 817, 113

\bibitem[{{Li} {et~al.}(2012){Li}, {Kong}, {Charles}, {Lu}, {Bartlett}, {Coe},
  {McBride}, {Rajoelimanana}, {Udalski}, {Masetti}, \& {Franzen}}]{Li2012}
{Li}, K.~L., {Kong}, A.~K.~H., {Charles}, P.~A., {et~al.} 2012, \apj, 761, 99

\bibitem[{{Liu}(2011)}]{Liu2011}
{Liu}, J. 2011, \apjs, 192, 10

\bibitem[{{Liu} \& {Di Stefano}(2008)}]{Liu2008}
{Liu}, J., \& {Di Stefano}, R. 2008, \apjl, 674, L73

\bibitem[{{Liu} {et~al.}(2013){Liu}, {Bregman}, {Bai}, {Justham}, \&
  {Crowther}}]{Liu2013}
{Liu}, J.-F., {Bregman}, J.~N., {Bai}, Y., {Justham}, S., \& {Crowther}, P.
  2013, \nat, 503, 500

\bibitem[{{Liu} {et~al.}(2015){Liu}, {Bai}, {Wang}, {Justham}, {Lu}, {Gu},
  {Liu}, {di Stefano}, {Guo}, {Cabrera-Lavers}, {{\'A}lvarez}, {Cao}, \&
  {Kulkarni}}]{Liu2015}
{Liu}, J.-F., {Bai}, Y., {Wang}, S., {et~al.} 2015, \nat, 528, 108

\bibitem[{{Mannucci} {et~al.}(2006){Mannucci}, {Della Valle}, \&
  {Panagia}}]{Mannucci2006}
{Mannucci}, F., {Della Valle}, M., \& {Panagia}, N. 2006, \mnras, 370, 773

\bibitem[{{Maravelias} {et~al.}(2014){Maravelias}, {Zezas}, {Antoniou}, \&
  {Hatzidimitriou}}]{Antoniou2014}
{Maravelias}, G., {Zezas}, A., {Antoniou}, V., \& {Hatzidimitriou}, D. 2014,
  \mnras, 438, 2005

\bibitem[{{Matson} {et~al.}(2015){Matson}, {Gies}, {Guo}, {Quinn}, {Buchhave},
  {Latham}, {Howell}, \& {Rowe}}]{Matson2015}
{Matson}, R.~A., {Gies}, D.~R., {Guo}, Z., {et~al.} 2015, \apj, 806, 155

\bibitem[{{McSwain} \& {Gies}(2005)}]{McSwain2005}
{McSwain}, M.~V., \& {Gies}, D.~R. 2005, \apjs, 161, 118

\bibitem[{{Meilland} {et~al.}(2012){Meilland}, {Millour}, {Kanaan}, {Stee},
  {Petrov}, {Hofmann}, {Natta}, \& {Perraut}}]{2012A&A...538A.110M}
{Meilland}, A., {Millour}, F., {Kanaan}, S., {et~al.} 2012, \aap, 538, A110

\bibitem[{{Morii} {et~al.}(2013){Morii}, {Tomida}, {Kimura}, {Suwa}, {Negoro},
  {Serino}, {Kennea}, {Page}, {Curran}, {Walter}, {Kuin}, {Pritchard},
  {Nakahira}, {Hiroi}, {Usui}, {Kawai}, {Osborne}, {Mihara}, {Sugizaki},
  {Gehrels}, {Kohama}, {Kotani}, {Matsuoka}, {Nakajima}, {Roming}, {Sakamoto},
  {Sugimori}, {Tsuboi}, {Tsunemi}, {Ueda}, {Ueno}, \& {Yoshida}}]{Morii2013}
{Morii}, M., {Tomida}, H., {Kimura}, M., {et~al.} 2013, \apj, 779, 118

\bibitem[{{Oemler} \& {Tinsley}(1979)}]{Oemler1979}
{Oemler}, Jr., A., \& {Tinsley}, B.~M. 1979, \aj, 84, 985

\bibitem[{{Oliveira} {et~al.}(2010){Oliveira}, {Steiner}, {Ricci}, {Menezes},
  \& {Borges}}]{Oliveira2010}
{Oliveira}, A.~S., {Steiner}, J.~E., {Ricci}, T.~V., {Menezes}, R.~B., \&
  {Borges}, B.~W. 2010, \aap, 517, L5

\bibitem[{{Orio}(2012)}]{Orio2012}
{Orio}, M. 2012, Bulletin of the Astronomical Society of India, 40, 333

\bibitem[{{Orio}(2013)}]{Orio2013}
---. 2013, The Astronomical Review, 8, 71

\bibitem[{{Orio} {et~al.}(2010){Orio}, {Nelson}, {Bianchini}, {Di Mille}, \&
  {Harbeck}}]{Orio2010}
{Orio}, M., {Nelson}, T., {Bianchini}, A., {Di Mille}, F., \& {Harbeck}, D.
  2010, \apj, 717, 739

\bibitem[{{Osborne}(2015)}]{Osborne2015}
{Osborne}, J.~P. 2015, Journal of High Energy Astrophysics, 7, 117

\bibitem[{{Panoglou} {et~al.}(2018){Panoglou}, {Faes}, {Carciofi}, {Okazaki},
  {Baade}, {Rivinius}, \& {Borges Fernandes}}]{Panoglou2018}
{Panoglou}, D., {Faes}, D.~M., {Carciofi}, A.~C., {et~al.} 2018, \mnras, 473,
  3039

\bibitem[{{Poeckert}(1979)}]{Poeckert1979}
{Poeckert}, R. 1979, \apjl, 233, L73

\bibitem[{{Porter} \& {Rivinius}(2003)}]{Porter2003}
{Porter}, J.~M., \& {Rivinius}, T. 2003, \pasp, 115, 1153

\bibitem[{{Raguzova}(2001)}]{Raguzova2001}
{Raguzova}, N.~V. 2001, \aap, 367, 848

\bibitem[{{Reig}(2011)}]{Reig2011}
{Reig}, P. 2011, \apss, 332, 1

\bibitem[{{Reig} {et~al.}(2016){Reig}, {Nersesian}, {Zezas}, {Gkouvelis}, \&
  {Coe}}]{Reig2016}
{Reig}, P., {Nersesian}, A., {Zezas}, A., {Gkouvelis}, L., \& {Coe}, M.~J.
  2016, \aap, 590, A122

\bibitem[{{Rivinius} {et~al.}(2013){Rivinius}, {Carciofi}, \&
  {Martayan}}]{2013A&ARv..21...69R}
{Rivinius}, T., {Carciofi}, A.~C., \& {Martayan}, C. 2013, \aapr, 21, 69

\bibitem[{{Sigut} \& {Jones}(2007)}]{Sigut2007}
{Sigut}, T.~A.~A., \& {Jones}, C.~E. 2007, \apj, 668, 481

\bibitem[{{Stanimirovic} {et~al.}(1999){Stanimirovic}, {Staveley-Smith},
  {Dickey}, {Sault}, \& {Snowden}}]{Stani1999}
{Stanimirovic}, S., {Staveley-Smith}, L., {Dickey}, J.~M., {Sault}, R.~J., \&
  {Snowden}, S.~L. 1999, \mnras, 302, 417

\bibitem[{{Struve}(1945)}]{Struve1945}
{Struve}, O. 1945, Popular Astronomy, 53, 259

\bibitem[{{Sturm} {et~al.}(2012){Sturm}, {Haberl}, {Pietsch}, {Coe},
  {Mereghetti}, {La Palombara}, {Owen}, \& {Udalski}}]{Sturm2012}
{Sturm}, R., {Haberl}, F., {Pietsch}, W., {et~al.} 2012, \aap, 537, A76

\bibitem[{{Sullivan} {et~al.}(2006){Sullivan}, {Le Borgne}, {Pritchet},
  {Hodsman}, {Neill}, {Howell}, {Carlberg}, {Astier}, {Aubourg}, {Balam},
  {Basa}, {Conley}, {Fabbro}, {Fouchez}, {Guy}, {Hook}, {Pain},
  {Palanque-Delabrouille}, {Perrett}, {Regnault}, {Rich}, {Taillet}, {Baumont},
  {Bronder}, {Ellis}, {Filiol}, {Lusset}, {Perlmutter}, {Ripoche}, \&
  {Tao}}]{Sullivan2006}
{Sullivan}, M., {Le Borgne}, D., {Pritchet}, C.~J., {et~al.} 2006, \apj, 648,
  868

\bibitem[{{Takei} {et~al.}(2008){Takei}, {Tsujimoto}, {Kitamoto}, {Morii},
  {Ebisawa}, {Maeda}, \& {Miller}}]{Takei2008}
{Takei}, D., {Tsujimoto}, M., {Kitamoto}, S., {et~al.} 2008, \pasj, 60, S231

\bibitem[{{Tammann}(1977)}]{Tammann1977}
{Tammann}, G.~A. 1977, in Astrophysics and Space Science Library, Vol.~66,
  Supernovae, ed. D.~N. {Schramm}, 95

\bibitem[{{Tammann}(1978)}]{Tammann1978}
{Tammann}, G.~A. 1978, \memsai, 49, 315

\bibitem[{{van den Heuvel} {et~al.}(1992){van den Heuvel}, {Bhattacharya},
  {Nomoto}, \& {Rappaport}}]{vandenHeuvel1992}
{van den Heuvel}, E.~P.~J., {Bhattacharya}, D., {Nomoto}, K., \& {Rappaport},
  S.~A. 1992, \aap, 262, 97

\bibitem[{{Wang} {et~al.}(2018){Wang}, {Gies}, \& {Peters}}]{Wang2018}
{Wang}, L., {Gies}, D.~R., \& {Peters}, G.~J. 2018, \apj, 853, 156

\bibitem[{{Waters} {et~al.}(1989){Waters}, {Pols}, {Hogeveen}, {Cote}, \& {van
  den Heuvel}}]{Waters1989}
{Waters}, L.~B.~F.~M., {Pols}, O.~R., {Hogeveen}, S.~J., {Cote}, J., \& {van
  den Heuvel}, E.~P.~J. 1989, \aap, 220, L1

\end{thebibliography}

\end{document}